\documentclass[sigconf]{acmart}

\usepackage{colortbl}
\usepackage{subfigure}
\usepackage[capitalise]{cleveref}
\usepackage{bm}
\usepackage{pifont}
\usepackage{marvosym}
\usepackage{stackengine}
\newcommand{\cmark}{\ding{51}}%
\newcommand{\xmark}{\ding{55}}%
\usepackage{multirow}

\author{Zhihua Zhong}
\authornote{Equal contribution.}
\email{zhongzhihua@zju.edu.cn}
\affiliation{%
\institution{State Key Lab of CAD\&CG, Zhejiang University}
\institution{Zhejiang University City College}
\country{China}}

\author{Jingsen Zhu}
\authornotemark[1]
\email{zhujingsen@zju.edu.cn}
\affiliation{%
\institution{State Key Lab of CAD\&CG, Zhejiang University}
\country{China}}

\author{Yuxin Dai}
\email{buttersdyx@gmail.com}
\affiliation{%
\institution{Zhejiang A\&F University}
\country{China}}

\author{Chuankun Zheng}
\email{ckzheng0320@zju.edu.cn}
\affiliation{%
\institution{State Key Lab of CAD\&CG, Zhejiang University}
\country{China}}

\author{Yuchi Huo}
\authornote{Corresponding author.}
\email{huo.yuchi.sc@gmail.com}
\affiliation{%
\institution{Zhejiang Lab}
\institution{State Key Lab of CAD\&CG, Zhejiang University}
\country{China}}

\author{Guanlin Chen}
\email{chenguanlin@zucc.edu.cn}
\affiliation{%
\institution{Zhejiang University City College}
\country{China}}

\author{Hujun Bao}
\email{bao@cad.zju.edu.cn}
\affiliation{%
\institution{State Key Lab of CAD\&CG, Zhejiang University}
\country{China}}

\author{Rui Wang}
\authornotemark[2]
\email{rwang@cad.zju.edu.cn}
\affiliation{%
\institution{State Key Lab of CAD\&CG, Zhejiang University}
\country{China}}

\setcitestyle{square}
\AtBeginDocument{%
  \providecommand\BibTeX{{%
    \normalfont B\kern-0.5em{\scshape i\kern-0.25em b}\kern-0.8em\TeX}}}

\copyrightyear{2023} 
\acmYear{2023} 
\setcopyright{acmlicensed}\acmConference[SA Conference Papers
'23]{SIGGRAPH Asia 2023 Conference Papers}{December 12--15, 2023}{Sydney,
NSW, Australia}
\acmBooktitle{SIGGRAPH Asia 2023 Conference Papers (SA Conference Papers
'23), December 12--15, 2023, Sydney, NSW, Australia}
\acmPrice{15.00}
\acmDOI{10.1145/3610548.3618209}
\acmISBN{979-8-4007-0315-7/23/12}

\begin{document}

\newcommand{\zjs}[1]{{\textcolor{magenta}{ZJS:#1}}}
\newcommand{\zck}[1]{{\textcolor[rgb]{0.2, 0.2, 0.8}{ZCK:#1}}}
\newcommand{\zzh}[1]{{\textcolor{red}{ZZH:#1}}}
\newcommand{\huo}[1]{{\textcolor{orange}{Huo:#1}}}
\newcommand{\new}[1]{{#1}}
\newcommand{\revision}[1]{{#1}}
\newcommand{\TBD}[1]{{\textcolor[rgb]{0.2,0.6,0.2}{#1}}}
\newcommand{\Skip}[1]{}

\def\d{\mathrm{d}}
\def\onedot{.~}
\def\eg{\emph{e.g}\onedot} \def\Eg{\emph{E.g}\onedot}
\def\ie{\emph{i.e}\onedot} \def\Ie{\emph{I.e}\onedot}
\def\cf{\emph{cf}\onedot} \def\Cf{\emph{Cf}\onedot}
\def\etc{\emph{etc}\onedot} \def\vs{\emph{vs}\onedot}
\def\wrt{w.r.t\onedot} \def\dof{d.o.f\onedot}
\def\iid{i.i.d\onedot} \def\wolog{w.l.o.g\onedot}
\def\etal{\emph{et al}\onedot}

\newcommand{\piHR}{\hat{I}^{HR}}
\newcommand{\iLR}{I^{LR}}
\newcommand{\gLR}{G^{LR}}
\newcommand{\gHR}{G^{HR}}
\newcommand{\iHIS}{I^{LR}_\mathrm{history}}

\newcommand{\bwi}{\bm{\omega}_i}
\newcommand{\bwo}{\bm{\omega}_o}

\definecolor{gold}{rgb}{0.945,0.898,0.675}
\definecolor{silver}{rgb}{0.882,0.882,0.882}
\definecolor{bronze}{rgb}{0.855,0.667,0.369}

\newcommand{\refFig}[1]{Figure \ref{#1}}
\newcommand{\refEq}[1]{Equation (\ref{#1})}
\newcommand{\refAlg}[1]{Algorithm \ref{#1}}
\newcommand{\refSup}[1]{Supplementary \ref{#1}}
\newcommand{\refSec}[1]{Section \ref{#1}}
\newcommand{\refTab}[1]{Table \ref{#1}}
\newcommand{\refApp}[1]{Appendix \ref{#1}}

\title{FuseSR: Super Resolution for Real-time Rendering through Efficient Multi-resolution Fusion}

\renewcommand{\shortauthors}{Zhihua Zhong*, Jingsen Zhu*, et al.}



\begin{abstract}
The workload of real-time rendering is steeply increasing as the demand for high resolution, high refresh rates, and high realism rises, overwhelming most graphics cards. To mitigate this problem, one of the most popular solutions is to render images at a low resolution to reduce rendering overhead, and then manage to accurately upsample the low-resolution rendered image to the target resolution, a.k.a. super-resolution techniques. 
\new{Most existing methods focus on exploiting information from low-resolution inputs, such as historical frames. The absence of high frequency details in those LR inputs makes them hard to recover fine details in their high-resolution predictions.
In this paper, we propose an efficient and effective super-resolution method that predicts high-quality upsampled reconstructions utilizing low-cost high-resolution auxiliary G-Buffers as additional input. 
With LR images and HR G-buffers as input, the network requires to align and fuse features at multi resolution levels. We introduce an efficient and effective H-Net architecture to solve this problem and significantly reduce rendering overhead without noticeable quality deterioration.
}
Experiments show that our method is able to produce temporally consistent reconstructions in $4 \times 4$ \new{and even challenging $8 \times 8$} upsampling cases at 4K resolution with real-time performance, with substantially improved quality \new{and significant performance boost} compared to existing works.{Project page: \url{https://isaac-paradox.github.io/FuseSR/}}
\Skip{Our method employs pre-integrated BRDF demodulation implemented on split-sum approximation to obtain high-resolution features at a negligibly low cost, making the reconstruction more reasonable and tractable. We further present a novel \emph{H-Net} architecture to losslessly fuse the high-resolution features into a low-resolution backbone using our dual-shuffling technique, which achieves an outstanding balance between quality and efficiency. 
Experiments show that our method is able to produce temporally consistent reconstructions in challenging $4 \times 4$ \new{and $8 \times 8$} upsampling cases at 4K resolution with real-time performance, with substantially improved quality compared to existing works \new{and significant performance boost compared to native HR rendering}.}
\end{abstract}

\ccsdesc[500]{Computing methodologies~Rendering}
\keywords{super resolution, rendering, deep learning}






\begin{teaserfigure}
\includegraphics[width=\linewidth]{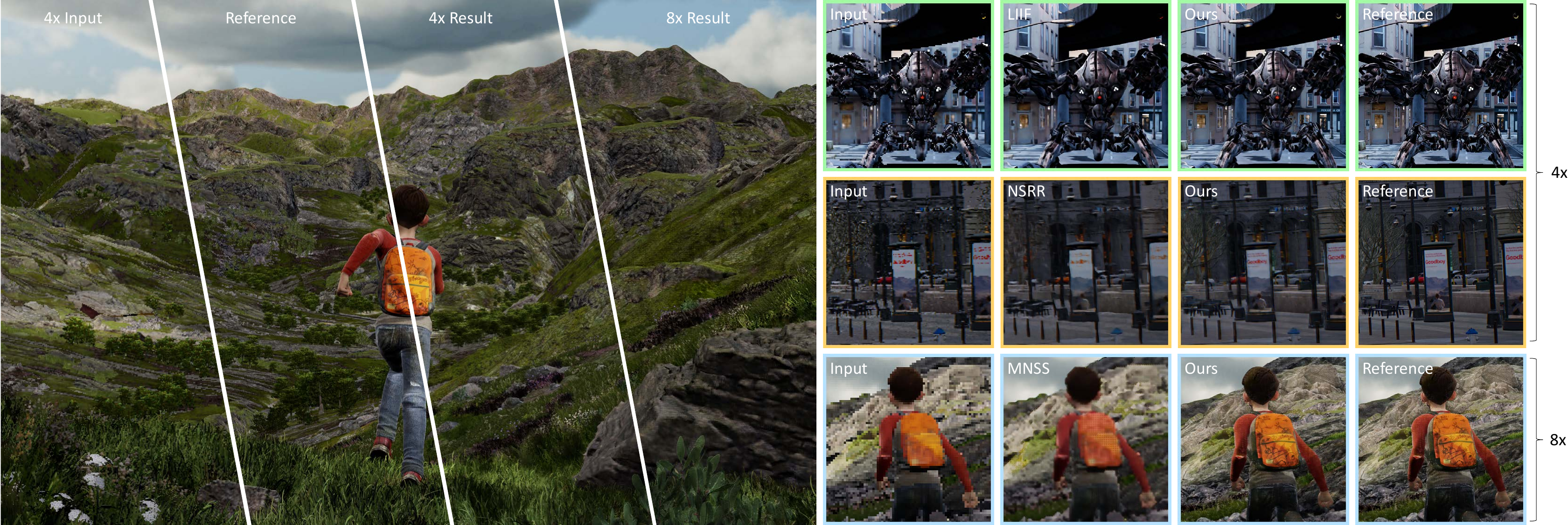}
\caption{\new{Our method achieves high-fidelity $4 \times 4$ super-resolution images, significantly outperforming existing methods in performance and quality. Our method even succeeds in the extremely challenging $8\times8$ super-resolution task (bottom right), on which existing methods basically fail, thanks to our BRDF demodulation and multi-resolution fusion design to preserve high-frequency details. Zoom in for details.}}
\end{teaserfigure}
\maketitle

\section{introduction}
In the past few years, with the popularity of high-resolution and high-refresh-rate displays, as well as \new{photographic realistic lighting and} the advancement of real-time ray tracing techniques, the computational workload of real-time rendering has increased dramatically. \new{The emergence of real-time raytracing has further increased the computational overhead of rendering for higher quality outputs.} 
Even with the high-end consumer GPU, rendering high-quality images at 144 FPS and 4K resolution is still extremely challenging. 
As a result, users have to make trade-offs between rendering quality, resolution, and refresh rate.

A large number of techniques have been proposed to alleviate this problem. Real-time denoising techniques \cite{fan2021real, schied2017spatiotemporal} render images at a low tracing budget and manage to reduce these images' noise levels to produce plausible results.
Frame extrapolation methods \cite{guo2021extranet, zeng2021temporally} focus on reconstructing accurate shading results from historical frames reprojected with motion vectors to accelerate rendering.
Foveated rendering methods \cite{kaplanyan2019deepfovea} propose to reduce the resolution at the periphery of users' vision without sacrificing perceived visual quality, thereby improving efficiency, but only for virtual reality headsets.

The most widely adopted and successful method is the super-resolution (SR) approaches, including DLSS \cite{dlss}, FSR \cite{fsr}, XeSS \cite{xess}, etc. Users can reduce the resolution of rendered images to decrease rendering time and upsample the low-resolution (LR) rendered image to obtain the final high-resolution (HR) image. 
However, they mainly consider upsampling factors less than $2\times 2$, which limits higher performance improvement. NSRR~\cite{xiao2020neural} pursues a more promising task that produces high-quality $4\times 4$-upsampled reconstruction in real-time by utilizing historical frames, yet it struggles with recovering accurate high-frequency texture details and cannot support real-time experience at resolutions higher than 1080p.

The results of NSRR demonstrate that high-resolution SR is a tough challenge. On the one hand, many high-fidelity details are lost even in historical frames. In theory, a $4\times 4$ SR reconstruction requires at least 16 historical frames to cover every pixel of the HR target fully, and such a long temporal window makes the history-reusing scheme basically infeasible in dynamic scenes. An intuitive solution is to utilize HR G-buffers that contain full-resolution information, of which the rendering cost is negligible and only sub-linearly increases. On the other hand, the network performance is also a critical concern for real-time SR and the neural network inference time increases rapidly w.r.t. the input resolution. Therefore, the need to increase feature resolution and reduce network bandwidth is a pair of contradictions that are difficult to resolve, which slows down the development of high-resolution SR.

In this paper, we present FuseSR, an \revision{efficient and effective} real-time super-resolution technique that is able to offer high-fidelity $4\times 4$ even $8 \times 8$ upsampled reconstruction with significantly improved quality and performance compared to existing works. Besides using historical information, we utilize HR G-buffer to provide per-pixel cues for the HR target. 
We further \Skip{follow the split-sum scheme to} decompose the shading results into pre-integrated BRDF and demodulated irradiance components, and train a network to predict the HR irradiance, to strike a better balance between quality and efficiency.
Most importantly, we propose \emph{H-Net} architecture to resolve the contradictions between HR features and LR bandwidth. \revision{In H-Net, we incorporate pixel shuffling and unshuffling~\cite{shi2016real,gharbi2016deep}} to losslessly align HR features with LR inputs and fuse the features into the LR network backbone, while preserving high-fidelity HR details. \Skip{Its novel dual-shuffling technique losslessly fuses the HR features into the LR network backbone and vice verse. Finally, we improve temporal stability with an attention-based history reusing module.} Our contributions can be summarized as follows:

\begin{itemize}
    \item 
    Our method successfully utilizes high-resolution G-buffers \Skip{in addition to historical information}to resolve the real-time super-resolution problem, which significantly outperforms existing methods in both \Skip{terms of}time and quality. \new{We are the first method to produce high-fidelity results in the challenging $8\times8$ super-resolution task.}
    
    \item 
    We propose H-Net, \revision{an efficient and effective} network design to conduct lossless multi-resolution feature alignment and fusion with a low-resolution network backbone. \revision{We innovatively employ pixel shuffling and unshuffling pair into our network design to align and fuse multi-resolution features into the same screen space.}
    
    \item 
    We introduce pre-integrated BRDF demodulation to resolve the super-resolution problem, improving detail preservation and reducing the redundancy of G-buffers.

    \Skip{\new{We introduce a simple yet effective \emph{channel-wise pooling} rather than traditional \emph{spatial pooling} to avoid losing high-frequency spatial details and align the correspondence between LR pixels and HR G-buffers.}}


\end{itemize}

\section{related work}

\paragraph{Real-time supersampling.}
In real-time rendering, each pixel of rendered images is point sampled. 
Therefore, the SR of rendered images can be conceived as a supersampling problem with upscaling.
Supersampling-based antialiasing techniques \cite{akeley1993reality, yong2006csaa} and temporal antialiasing methods \cite{karis2014, yang2020survey} manage to conduct supersampling in spatial and temporal domain.
\Skip{Supersampling-based antialiasing techniques \cite{akeley1993reality, yong2006csaa} focus on performing multiple sampling in undersampled pixels to obtain smooth edges.
Temporal antialiasing methods \cite{karis2014, yang2020survey} extend supersampling from spatial domain to temporal domain.
These methods manage to apportion supersampling to the rendering of different frames by reusing the valid samples in the history frames.}
These aforementioned antialiasing techniques supersample without resolution changing.
On the other hand, deep learning-based supersampling with upscaling has gained increasing attention recently.
Deep learning super sampling (DLSS) \cite{dlss} utilizes neural networks to upscale LR frames, significantly reducing the rendering cost of HR frames and enabling high-quality rendering at high resolution in real-time.
It inspired a series of similar works, including FSR \cite{fsr}, TAAU \cite{taau}, and XeSS \cite{xess}.
However, these methods focus on tasks with upsampling factors smaller than $2 \times 2$, limiting further performance improvement.
NSRR \cite{xiao2020neural} is the closest to our method, which uses temporal dynamics and G-buffer to provide compelling results in the challenging 4$\times$4 upsampling case, but it cannot accurately recover HR details and is unable to support real-time rendering at high resolutions such as 2K and 4K. \new{MNSS~\cite{MNSS} also leverages historical frames, and achieves competitive runtime performance due to its lightweight network design. }


\paragraph{\revision{Pixel Shuffling.}}
\revision{ESPCN~\cite{shi2016real} firstly designs pixel shuffling as an efficient upscaling operation in their super-resolution network, which keeps most convolutional layers in low resolution to reduce network computational overhead. \citet{gharbi2016deep} further design a pixel unshuffling-shuffling pair in their denoising network architecture, with the purpose of converting pixel-wise noise patterns into channel-wise for better denoising processing. Inspired by these methods, our approach adopt pixel shuffling for high speed and pixel unshuffling for the alignment and fusion of multi-resolution input features.
}


\paragraph{BRDF demodulation.}
Materials with fine details can increase rendering realism but also make reconstruction tasks (e.g., denoising and supersampling) more challenging.
Demodulating BRDF is a common practice for preserving details in reconstruction tasks. \citet{bako2017kernel} and SVGF \cite{schied2017spatiotemporal} demodulate the diffuse albedo from the noisy image and then modulate it with the denoised result to alleviate the blurring problem.
Guo et al. \shortcite{guo2021extranet} found that such demodulation was also beneficial for extrapolated rendering.
\Skip{Demodulating unshadowed direct illumination \cite{heitz2018combining} is adopted for the computation of soft shadow.}
Pre-integrated BRDF demodulation \cite{zhuang2021real} is also proven effective for real-time denoising.
Inspired by these methods, our approach introduces demodulation similar to \cite{zhuang2021real} to decompose the color frames SR into the SR tasks of corresponding pre-integrated BRDF maps and demodulated irradiance maps, wherein the demodulated irradiance map tends to be smoother than the original color, thus reducing the SR difficulty. 
\Skip{The pre-integrated BRDF map can be precomputed and efficiently acquired as a G-Buffer at the target resolution. Benefiting from this demodulation, our approach is able to produce better SR predictions with substantially improved detail preservation compared to other advanced works.}

\begin{figure*}[ht!]
    \centering
    \includegraphics[width=\textwidth]{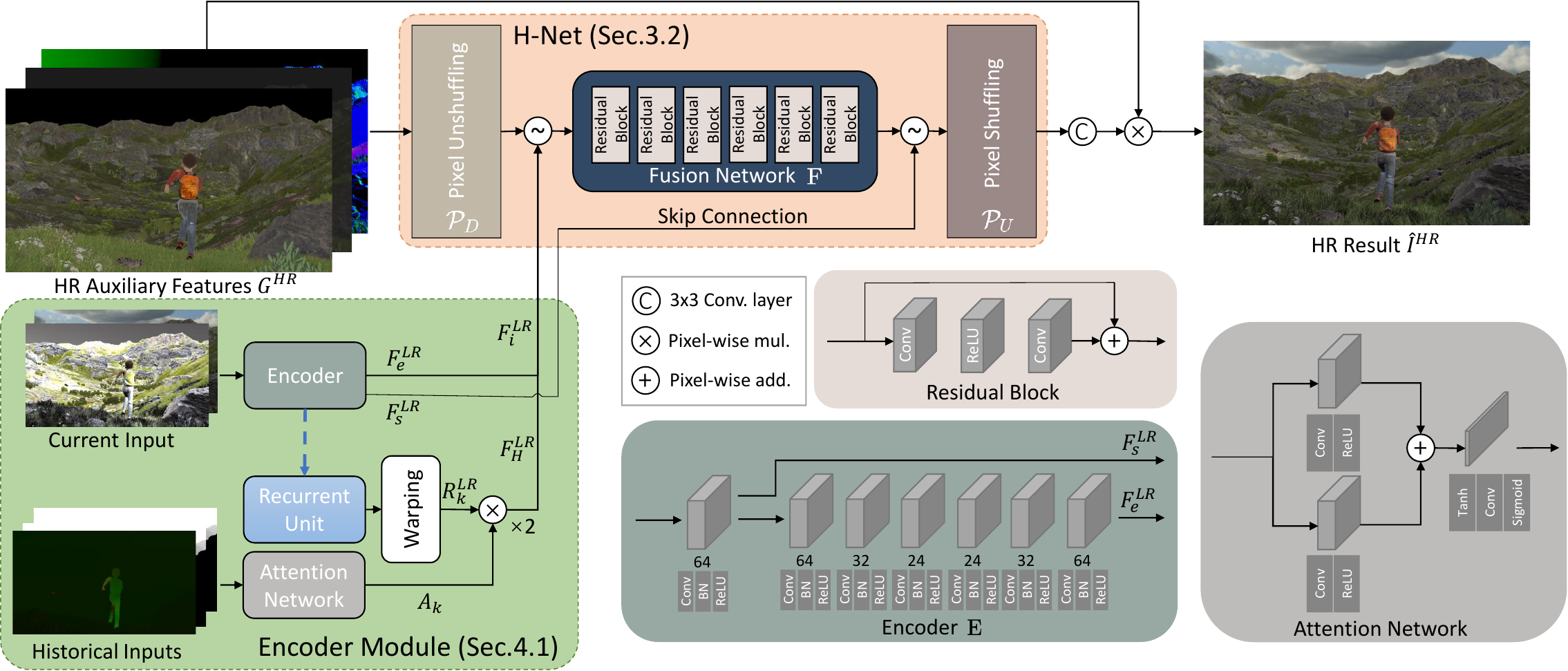}
    \caption{The pipeline of our method. We perform BRDF demodulation to turn the SR of shading images into the prediction of the HR demodulated irradiance map and acquisition of the HR pre-integrated BRDF map. An effective and efficient \emph{H-net} is introduced to fuse LR features with HR G-buffers to reconstruct the HR demodulated irradiance map and remodulate it with the HR pre-integrated BRDF map that can be acquired efficiently to obtain the final SR outcome. \Skip{We further employ historical inputs in the encoder module to improve temporal stability (see supplementary).} }
    \label{fig:architectire}
\end{figure*}

\section{Method}
\label{sec:method}

The goal of our work is to reconstruct upsampled HR frames $\piHR$ from the corresponding LR frames $\iLR$ in real-time rendering.
Unfortunately, the SR problem that relied entirely on LR color frames is ill-posed because of the lack of HR details.
Inspired by previous works \cite{guo2021extranet}, we take HR G-buffers as SR cues to make the problem more tractable.
G-buffer is a commonly used rendering byproduct containing rich scene information (e.g., depths, normals, and texture details), 
and its acquisition is significantly cheaper (several milliseconds for each 1080p frame) than heavy shading and potential post-processing tasks.

To utilize material information from G-Buffer, we employ pre-integrated BRDF demodulation \Skip{implemented based on G-buffers} (\cref{sec:demod}) to explicitly filter out HR details, turning the color frame SR to an easier demodulated irradiance SR problem. 
\new{We further design our \emph{H-Net} architecture employing pixel shuffling operations (\cref{sec:shuf}) to effectively and efficiently align and fuse HR and LR information.} Our design demonstrates significantly improved quality and performance compared with other advanced works.

To be precise, we describe our task as follows:
\begin{equation}
    \piHR = \textbf{SuperResolution}(\iLR, \gLR, \gHR),
    \label{equ:sr_task}
\end{equation}
\new{where $\gLR$ and $\gHR$ denote LR and HR G-buffers respectively, which are taken as auxiliary input features to enrich the input information for better predictions.}

\Skip{
In this section, we introduce the problem setting of our superresolution task (\cref{sec:overview}) and describe our learning-based framework in detail.
As the overview of our method outlined in \cref{fig:architectire},
we first turn the color frame SR to the easier demodulated illuminated SR problem by BRDF demodulation (\cref{sec:demod}) and then utilize a novel dual shuffling design to obtain high-quality SR results with high efficiency (\cref{sec:shuf}).
For clarity, the specific network architecture is not covered in this section, and we will explain it in the next section.

\subsection{Problem Setting}
\label{sec:overview}
Our work aims to generate $4\times4$-upsampled HR frames from the corresponding LR frames for real-time rendering. In addition to LR color images, our method takes advantage of auxiliary information commonly available in modern game engines for superresolution. Specifically, we obtain an HR frame $\piHR$ from its LR version $\iLR$, LR G-buffers $\gLR$, and HR G-buffers $\gHR$:
\begin{equation}
    \piHR = \textbf{SuperResolution}(\iLR, \gLR, \gHR).
    \label{equ:sr_task}
\end{equation}

\paragraph{High-resolution G-buffers.} G-buffers are common intermediate results in deferred rendering, which contain rich scene information such as texture details, normals, and depths. 
Generally, G-buffer acquisition is significantly cheaper (several milliseconds for each 1080p frame) 
than heavy shading and potential post-processing tasks, allowing us to provide (even HR) G-buffers as cues for super-resolution at a low cost.
High-resolution G-buffers can provide additional information (e.g., HR texture details and geometry edges) that is unavailable in LR versions and valuable for super-resolution.
Therefore, our method takes both LR and HR G-buffers as input to enhance the quality. 
In our experiments, the LR G-buffers contain depths, normals, and motion vectors, and the HR G-buffers are comprised of material roughness and view directions. \zck{Ensure these attributes are correct.}
}

\subsection{BRDF Demodulation for Superresolution}
\label{sec:demod}

Today's production-ready 3D scenes are becoming more and more exquisite, making the superresolution task, especially those with high-frequency features, more challenging.
Inspired by \citet{zhuang2021real}, we perform BRDF demodulation to filter out high-frequency material details for better overall quality and detail preservation.
Specifically, we reformulate the rendering equation~\cite{kajiya1986rendering} into the multiplication of the pre-integrated BRDF term $F_\beta(\bwo)$ and the demodulated irradiance term $L_D(\bwo)$:
\begin{align}
    L_o(\bwo)&=\int_{\Omega}f_r(\bwi,\bwo)L_i(\bwi)\cos\theta_i\d\bwi \\
    F_\beta(\bwo)&=\int_{\Omega}f_r(\bwi,\bwo)\cos\theta_i\d\bwi,\ L_D(\bwo) = \frac{L_o(\bwo)}{F_\beta(\bwo)} 
\end{align}
where $\bwi, \bwo$ denotes the incoming and outgoing directions, respectively, $f_r(\bwi,\bwo)$ represents the bidirectional reflectance distribution function (BRDF) at the shading point, the lighting function $L_i(\bwi)$ describes incident radiance at the shading point,  $\cos\theta_i$ is the cosine term, and $L_o(\bwo)$ is the outgoing radiance.
This demodulation decomposes the radiance of a pixel into two terms: $F_\beta(\bwo)$ and $L_D(\bwo)$.
Hence the superresolution task turns into estimating the corresponding HR $F_\beta$ and HR $L_D$ maps.

We adopt the precomputation approach proposed by Karis et al. \shortcite{karis2013} (known as \emph{split-sum approximation}) to acquire $F_\beta$ maps at a negligibly low cost.
In the precomputation stage, the BRDF integrals to each normal and light direction combination on varying roughness values are computed and stored in a 2D lookup texture (LUT).
After that, we can easily obtain $F_\beta$ maps with arbitrary resolutions by querying the LUT according to each pixel's\Skip{material} roughness and view direction\Skip{attributes} that are provided in the G-buffers.
We refer readers to the original paper \cite{karis2013} for more details.

Then, we leverage a neural network $\Phi$ to predict the HR demodulated irradiance map $\hat{L}_D^{HR}$ and multiply it with the HR pre-integrated BRDF map $F_\beta^{HR}$ pixel-by-pixel to obtain the SR outcome. 
Thus \cref{equ:sr_task} can be rewritten as:
\begin{equation}
\label{equ:demod}
    \hat{L}_D^{HR} = \Phi(L_D^{LR}, \gLR, \gHR),\ \piHR = F_\beta^{HR} \odot \hat{L}_D^{HR},
\end{equation}
where $\odot$ denotes pixel-wise multiplication. The LR demodulated irradiance map $L_D^{LR}$ can be easily computed by dividing the LR color map $\iLR$ by the LR pre-integrated BRDF map $F_\beta^{LR}$ pixel-by-pixel, while
$F_\beta^{HR}$ can be easily pre-computed from HR G-buffers $\gHR$.
Note that we omit the emitted radiance term in the rendering equation because its HR version is available in the G-buffers.

\revision{Please refer to our supplementary for the visualization of $F_\beta$ map with high-frequency details and $L_D$ map smoother than the original pixel color.} 
We believe that estimating the smoother $L_D$ map instead of the original color frame is conducive to improving the quality and generalizability. Our experiments confirmed this choice (see \cref{sec:ab_module}). 





\subsection{\new{H-Net: An Effective and Efficient Network for Multi-Resolution Alignment and Fusion}}
\label{sec:shuf}

{
Given that our network takes multi-resolution features (LR image and HR G-buffers) as input, to make full use of the HR auxiliary inputs, an efficient network is required to organize the features at different resolution levels. Aligning pixels that share the same screen space position can preserve correct spatial correlation between multi-resolution input features. After that, our network fuses the aligned multi-resolution features to finish the SR task.

\begin{figure}[ht]
    \centering
    \includegraphics[width = 0.8\linewidth]{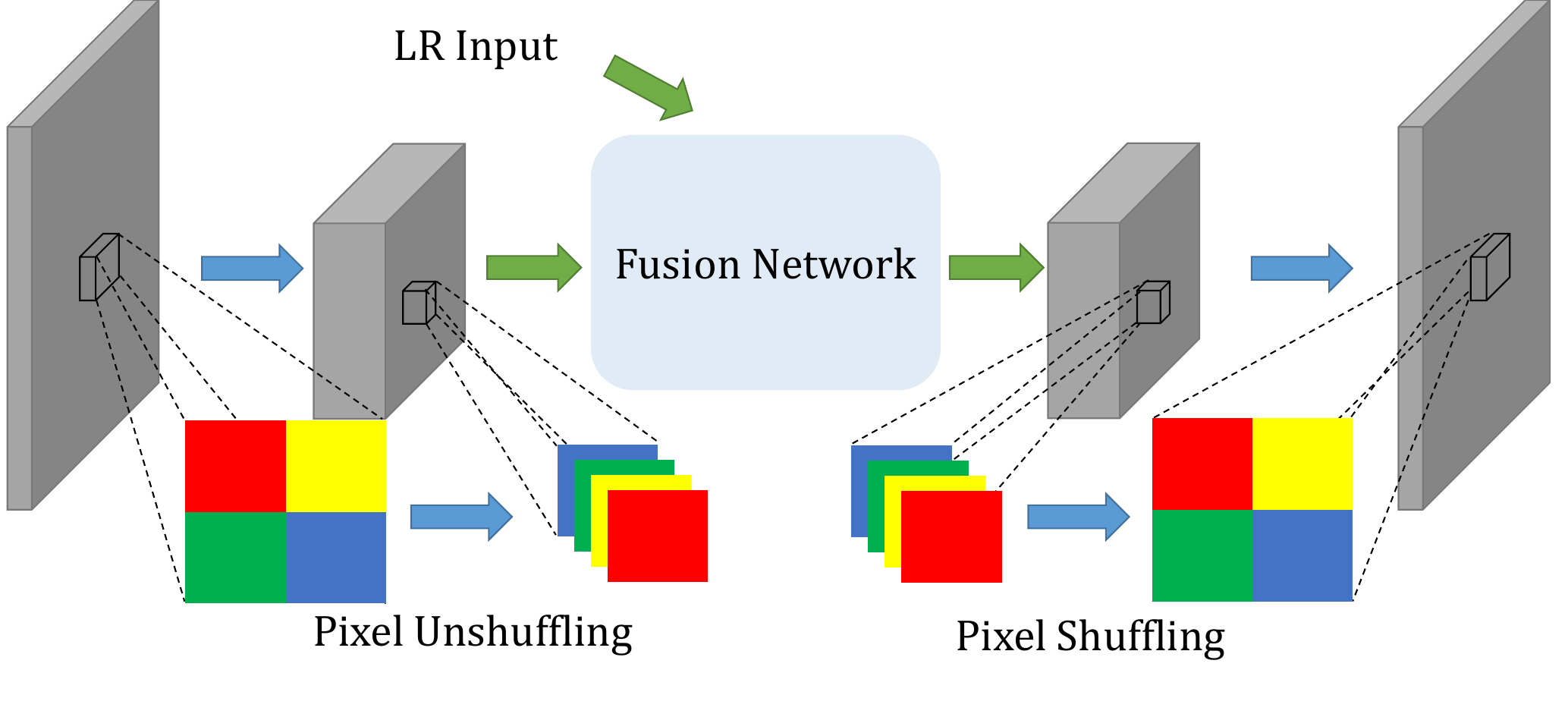}
    \caption{Architecture of H-Net.}
    \label{fig:dual_shuffling}
\end{figure}

\paragraph{Multi-resolution feature alignment.} To align pixels sharing the same screen space position, \textit{upsampling} and \textit{pooling} are two common strategies, where the former aligns at HR level and the latter aligns at LR level. Considering that the performance of convolutional networks degrades sharply as the input resolution increases, the upsampling alignment contradicts with the need of real-time performance, while pooling is a more feasible choice.

Pooling operations are commonly used in neural networks, including maximum pooling and average pooling. However, these pooling operations inevitably damage the spatial details such as HR edges and textures, which are key information for SR quality improvements. \revision{Therefore, instead of lossy pooling operations, we adopt pixel unshuffling operation~\cite{shi2016real,gharbi2016deep} to align an HR feature to LR without information loss.} We find that pixel unshuffling can losslessly shrink an HR feature map into LR space, converting pixel-wise spatial information into channel-wise deep information. Specifically, this operation divides the HR map into blocks of size $r \times r$ ($r$ is the downscaling factor) and concatenates the features of all pixels within each block to form pixels of the LR version, \ie transforming a map of shape $[C, H*r, W*r]$ into an LR map of shape $[C*r*r, H, W]$ without information loss.

 
\paragraph{H-Net} We propose an \textit{H-Net} architecture with a pixel unshuffling-shuffling pair, to efficiently align and fuse LR and HR information to faithfully preserve HR details during the alignment and fusion. The architecture of H-Net is shown in \cref{fig:dual_shuffling}:
\begin{enumerate}
    \item We use pixel unshuffling $\mathcal{P}_D$ to downscale $\gHR$ and concatenate it with other LR inputs.
    \item The concatenated feature is fed into a fusion network backbone $\mathbf{F}$. Note that $\mathbf{F}$ runs at LR level, which prevents significant HR computational overhead.
    \item The output of $\mathbf{F}$ is transformed to HR space using pixel shuffling $\mathcal{P}_U$ to obtain the HR output $\hat{L}_D^{HR}$.
\end{enumerate}

}

\Skip{
\new{A straightforward network architecture to fuse HR and LR features is to upsample the LR information and concatenate it with the HR information, as the previous work \cite{kim2016accurate} does. However, the performance of convolutional networks degrades sharply as the input resolution increases. The incorporation of HR inputs inherently contradicts with the original intention of framerate boosting of SR tasks, so the network needs to keep } 
\Skip{A straightforward alternative is to upsample the LR information and concatenate it with the HR information, as the previous work \cite{kim2016accurate} does. 
The main problem of this scheme is the redundancy of LR information and high computational cost}
Another intuitive solution is to downsample the HR map into LR using pooling operations, but this operation inevitably damages the fine details in HR that is crucial for quality improvements in SR. \new{How to design a network architecture to effectively and efficiently fuse HR information is the key challenge to resolve this contradiction.}

To address this challenge, we propose \emph{H-Net}, a novel network architecture using \emph{dual shuffling} for feature fusion, to make full use of both LR and HR information while being efficient.
As shown in \cref{fig:dual_shuffling}, H-Net mainly consists of three parts: a fusion LR backbone $\mathbf{F}$, a pixel shuffle-based upscaling module $\mathcal{P}_U$, and a pixel shuffle-based downscaling module $\mathcal{P}_D$ (where the two pixel shuffle modules are collectively referred to as \emph{dual shuffling}).

{
\color{cyan}
\begin{enumerate}
    \item We utilize the downscaling module $\mathcal{P}_D$ to reduce the resolution of the HR information to the same size as the LR information. $\mathcal{P}_D$ is implemented as the inverse of pixel shuffle operation \cite{shi2016real}, which benefits from the avoidance of HR detail losses during resolution reduction.
This inverse operation divides one map into blocks of size $r \times r$ ($r=4$ in our case) and concatenates the features of all pixels within each block to form pixels of the LR version, \ie transforming a map of shape $[C, H*r, W*r]$ into an LR map of shape $[C*r*r, H, W]$ losslessly.
    \item The transformed $F_\beta^{HR}$ is concatenated in channel dimension with the LR feature map obtained by encoding the LR demodulated irradiance map $L_D^{LR}$ and LR G-buffers $\gLR$. The concatenated feature map is then fed into the fusion network $\mathbf{F}$ to obtain the output with information fused.
    \item The output of fusion network is transformed to the HR space using the subsequent pixel shuffle-based upscaling module $\mathcal{P}_U$ to obtain the HR demodulated irradiance map $\hat{L}_D^{HR}$.
\end{enumerate}
Note that the feature map is kept consistently at low resolution throughout the fusion network to prevent significant computational overhead.
}

\Skip{Firstly, we utilize the downscaling module $\mathcal{P}_D$ to reduce the resolution of the HR information to the same size as the LR information.
To avoid loss of HR details in the resolution reduction, we implement the inverse of pixel shuffle operation \cite{shi2016real} as our downscaling module.
This inverse operation divides one map into blocks of size $r \times r$ ($r=4$ in our case) and concatenates the features of all pixels within each block to form pixels of the LR version, \ie transforming a map of shape $[C, H*r, W*r]$ into an LR map of shape $[C*r*r, H, W]$ losslessly.
Next, the transformed $F_\beta^{HR}$ is concatenated in channel dimension with the LR feature map obtained by encoding the LR demodulated irradiance map $L_D^{LR}$ and LR G-buffers $\gLR$. Last, the concatenated feature map is fed into the fusion network $\mathbf{F}$ and the output is then transformed to the HR space using the subsequent pixel shuffle-based upscaling module $\mathcal{P}_U$ to obtain the HR demodulated irradiance map $\hat{L}_D^{HR}$.
Note that the feature map is kept consistently at low resolution throughout the fusion network to prevent significant computational overhead. }
}

Formally, the process can be expressed as follows:
\begin{equation}
    \hat{L}_D^{HR} = \mathcal{P}_U\left(\mathbf{F}\left(\left[\mathbf{E}\left(\left[L_D^{LR}, \gLR\right]\right), \mathcal{P}_D\left(\gHR\right)\right]\right)\right), 
    \label{equ:dual}
\end{equation}
where $[\cdot,\cdot]$ represents concatenation, and $\mathbf{E}$ is the encoder used to obtain the LR feature map (\cref{sec:network}).
The name ``H-Net'' is motivated by the shape of the network with two HR ends and an LR bottleneck, resembling an ``H''.

With the use of pixel unshuffling, our method can faithfully preserve the HR details, thus facilitating high-quality SR outcomes.
Moreover, since neighboring pixels are usually highly correlated, aggregating neighboring pixels of the HR map together via $\mathcal{P}_D$ allows obtaining more compact implicit representations in the following fusion network, which is beneficial for reducing data redundancy and enhancing efficiency. \new{Our experiments show that our alignment strategy even achieves better quality than the na\"ive HR upsampling alignment strategy (see \cref{sec:ab_align}).}

\Skip{
Existing super-resolution works~\cite{xiao2020neural}\zjs{Add cite: XeSS, DLSS} also exploit historical frames to provide temporal information for the network so as to enhance the prediction quality. In our method, despite the fact that the utilization of HR features already yields sufficient quality improvement, we can still leverage historical frames to strive for further refinement. Considering the balance between quality and performance (\zjs{ref: performance table}), we leave this as an optional module of our network. The historical frames are reprojected onto current screen space and encoded as features into the network.

Nevertheless, the reprojected historical frames are not reliable everywhere due to the changes of dynamic scene and occlusion relationships, which can be visualized as ghosting and shimmering. In order to eliminate these artifacts, we design an attention-based network to localize and mask out unreliable regions in historical frames. We adopt the design of additive attention~\cite{bahdanau2014neural}:
\begin{equation}
    \label{equ:attention}
    \begin{split}
        H &= \mathrm{concatenate}(M,D_\mathrm{current},D_\mathrm{history}) \\
        A &= \sigma(\textbf{Conv}_\mathrm{v}(\tanh(\textbf{Conv}_\mathrm{a}(H)+\textbf{Conv}_\mathrm{b}(H))))
    \end{split}
\end{equation}
where $\sigma$ is the sigmoid activation function, $M$ is the motion vector, and $D_\mathrm{current},D_\mathrm{history}$ are depth maps of current and historical frames, respectively. The attention network generates an attention map $A$ that emphasizes regions of interest and masks out unreliable regions. $A$ will be further pixel-wise multiplied by the encoded historical feature:
\begin{equation}\label{equ:mask}
    F_\mathrm{masked} = A \odot \textbf{Encode}(I_\mathrm{history})
\end{equation}
The masked encoded feature $F_\mathrm{masked}$ will be concatenated with \zjs{TBD}
}

\Skip{
Modern game engines provide ready-made historical information including frame colors, depth maps and motion vectors, which are utilized by our model to provide temporal information for superresolution task. The historical frames are reprojected onto the current screen space according to the motion vectors. However, it is inevitable for a temporal method that reuse historical information to handle the differences brought by the changes of virtual scene, which can be visualized as ghosting and shimmering. \zjs{To be refined}

In order to eliminate the influence of these artifacts, TAA~\cite{yang2020survey} leverages different heuristic methods to rectify the reprojected historical data according to the samples of current frame; 
NSRR~\cite{xiao2020neural} designs a reweighting module which generates a confidence map to emphasize the unblocked area;
\citet{guo2021extranet} heuristically computes the occlusion motion vectors~\cite{zeng2021temporally} to compensate for the occluded background in historical frames.

All of these methods can be categorized into two basic approaches: rectification and occluded area localization. For learning-based methods, the networks after the concatenation of current frame and historical frames are responsible for rectification and temporal information blending, where historical features are selectively used to enhance the feature information of current frame.

\begin{figure}
    \centering
\subfigure[Current frame]{
\includegraphics[width = 0.30\linewidth]{images/placeholder.png}}
\subfigure[Warped history frame]{
\includegraphics[width = 0.30\linewidth]{images/placeholder.png}}
\subfigure[History frame]{
\includegraphics[width = 0.30\linewidth]{images/placeholder.png}}\\
\subfigure[Attention map]{
\includegraphics[width = 0.30\linewidth]{images/placeholder.png}}
\subfigure[Difference of current frame and warping result]{
\includegraphics[width = 0.30\linewidth]{images/placeholder.png}}
\subfigure[Second moment norm of motion vector]{
\includegraphics[width = 0.30\linewidth]{images/placeholder.png}}

    \caption{Attention map}
    \label{fig:attention_map}
\end{figure}

On the other hand, the occluded areas in historical frames can be located according to the motion vectors and depth information. We design a CNN-based attention network which generates an attention map, as visualized in \cref{fig:attention_map}, to mask out the occluded areas with soft attention values, helping the network ignore unreliable historical information. Our attention network adopts the design of additive attention~\cite{bahdanau2014neural}, replacing the fully connected layers in the original design with a single-layer convolution network, as formulated in \cref{equ:attention}. The attention network takes as input the motion vectors, along with the depth maps for both current and historical frames, and outputs an attention map that emphasizes areas of interest and masks out unreliable areas:

\begin{equation}
    \label{equ:attention}
    \begin{split}
        H &= \mathrm{concatenate}(M,D_\mathrm{current},D_\mathrm{history}) \\
        A &= \sigma(\textbf{Conv}_\mathrm{v}(\tanh(\textbf{Conv}_\mathrm{a}(H)+\textbf{Conv}_\mathrm{b}(H))))
    \end{split}
\end{equation}
where $M$ is the motion vector and $D_\mathrm{current},D_\mathrm{history}$ are depth maps for current and historical frames respectively, and $\sigma$ is the sigmoid activation function which limits the attention value within $(0,1)$. The attention map $A$ will further be multiplied to the feature map encoded from historical frames:
\begin{equation}\label{equ:mask}
    F_\mathrm{masked} = A \otimes \textbf{Encode}(I_\mathrm{history})
\end{equation}
where $\otimes$ denotes pixel-wise tensor multiplication, and $\textbf{Encode}$ is the encoding operation defined later in \cref{equ:encode}.
}



\Skip{NSRR~\cite{xiao2020neural} encodes historical frames into historical features to recover the details of current frame, which requires additional convolutional computations and brings inevitable costs in real-time rendering.
Most of the previous works use different encoder designs or different color spaces between current and historical frames, requiring at least two encoding passes to separately process with current and historical frames. However, the encoding computations of historical frames are redundant, since historical frame shares the same space as current frame and its encoding process should be no different from current frame's. Thus, the encoded historical features can be reused to save encoding time.}


\Skip{Based on this observation, we design a feature reuse mechanism, as shown in \cref{fig:feature_reuse}, to further reduce the computation time. Our feature reuse mechanism uses a shared the encoder network module for both current and historical frames. When applied to real-time rendering, the encoded features of current frame will be cached and reused as historical features in the future. Given motion vectors, historical features can be projected onto current screen space directly at runtime. This mechanism saves the encoding time of historical features, bringing significant performance optimizations.
}

\section{Network and Training}
\label{sec:network}

In this section, we first provide the specific architecture of our network in \cref{sec:arch}, and then describe details of the implementation and training in \cref{sec:train}.

\subsection{Network Architecture}
\label{sec:arch}
Here we explain the architectural details of each component involved in \cref{equ:dual}, including the encoder $\mathbf{E}$, fusion network $\mathbf{F}$, and a pixel unshuffling-shuffling pair $\mathcal{P}_D$ and $\mathcal{P}_U$.
Unless otherwise stated, all convolutional layers are set by default to a convolution with a $3 \times 3$ kernel size, a stride of 1, zero padding, and a subsequent ReLU activation. Please refer to our supplementary material for detailed network configurations.

\paragraph{Encoder} The encoder module takes LR irradiance and G-buffers as input and outputs an LR feature map as part of the input of our H-Net (\cref{equ:dual}). Following existing works \Skip{\cite{xiao2020neural,dlss,fsr,xess}}, we also utilize LR historical frames to encourage temporal consistency. \new{The architecture of the encoder is shown in the bottom left corner of \cref{fig:architectire}, and please refer to our supplementary material for details on how we encode LR current frame and historical frames.}

Formally, the encoder takes the LR irradiance along with G-buffers of current frame and 2 previous frames as input, and output 2 LR feature maps $F_s^{LR}$ and $F_i^{LR}$:
\begin{equation}
    F_s^{LR}, F_i^{LR} = \mathbf{E}\left( \left[ L^{LR}_{D_i}, G^{LR}_i, \left\{ L^{LR}_{D_k}, G^{LR}_k \right\}_{k\in\{i-1,i-2\}} \right] \right)
\end{equation}
where $i$ is the current frame index, $F_s^{LR}$ is the output of the first layer of $\mathbf{E}$ used for skip connection later, and $F_i^{LR}$ is the final output as an input of the fusion network.

\Skip{\paragraph{Encoder}
We design the architecture of the encoder network $\mathbf{E}$ as a 7-layer convolutional network, which takes the LR depth map, normal map, and demodulated irradiance as input and outputs an LR feature map with 32 channels.
The hidden channels of each layer in $\mathbf{E}$ are set to 64, 64, 32, 24, 24, and 32, respectively.
We utilize $\mathbf{E}$ to encode each frame, obtaining the corresponding feature map $F_i^{LR}$ where $i$ is the frame index.}

\Skip{\paragraph{Encoder and historical re-using module.} Inspired by existing works \cite{xiao2020neural,dlss,fsr,xess}, we employ a history-reusing mechanism to encourage temporal consistency.
Specifically, we encode the past two frames into feature maps $F_k^{LR}, k \in \{i-1, i-2\}$ ($i$ represents the current frame's index), and reproject them to the current view according to the motion vectors.
However, the reprojected feature maps $R_k^{LR}$ may be unreliable in some pixels due to the changes in dynamic scene and occlusion relationships, which can cause notorious artifacts such as ghosting and shimmering.
In order to mitigate these artifacts, we exploit an attention-based network to localize and mask out unreliable regions in the reprojected history frames. 
As shown in \cref{fig:architectire}, we adopt the design of additive attention \cite{bahdanau2014neural} to compute an attention map $A_k$ for each history frame individually:
\begin{equation}
    \begin{split}
        H &= \left[D_i^{LR}, D_k^{LR}, M_k^{LR} \right], k \in \{i-1, i-2\}\\
        A_k &= \mathrm{sigmoid}(\mathbf{Conv_v}(\tanh(\mathbf{Conv_a}(H)+\mathbf{Conv_b}(H)))),
    \end{split}
    \label{equ:attention}
\end{equation}
where $D_k^{LR}$ are the depth maps of the $k$-th frame,
$M_k^{LR}$ is the motion vector, $[\cdot]$ represents concatenation in channel dimension,
and $\mathbf{Conv_*}$ represent a single-layer convolution.
We pixel-wise multiply the attention map $A_k$ with $R_k^{LR}$ to filter out invalid pixels.
Then, the two filtered feature maps are concatenated, forming a history feature map $F_H^{LR}$ to be used in the following fusion network.}

\Skip{
\subsubsection{Encoder}
We design the architecture of the encoder network $\mathbf{E}$ as a 7-layer convolutional network, which takes LR G-buffers $\gLR$ (comprised of depths and normals) and demodulated irradiance $L_D^{LR}$ as input. The complete encoder $\mathbf{E}$ is divided into 2 sub-networks:
\[
    E = E_1 \circ E_0 \label{equ:encode}
\]
The first convolutional layer of the encoder ($E_0$) yields a 64-channel primary feature $F_\mathrm{primary}$, which will be skip-connected to the last stage of the network to retain primary input information. The subsequent 6 layers of convolution network ($E_1$) will further extract a 32-channel feature $F_\mathrm{current}$ from the shallow primary feature:
\begin{align}
    F_\mathrm{primary} &= E_0([L_D^{LR}, D^{LR}, N^{LR}]) \\
    F_\mathrm{current} &= E_1(F_\mathrm{primary}) = E([L_D^{LR}, D^{LR}, N^{LR}])
\end{align}
The numbers of hidden channels within $E_1$ is 64, 32, 24, 24, 32 respectively, \TBD{where we empirically find a decreased hidden unit beneficial}.
}

\paragraph{Fusion network}
In our fusion network $\mathbf{F}$, the LR feature from encoder $F_i^{LR}$, and the unshuffled HR auxiliary features $\gHR$\Skip{(mentioned in \cref{sec:shuf})} are concatenated and fed into a subsequent network $\mathbf{F}$ to obtain an LR fused feature map $F_f^{LR}$:
\begin{equation}
    F_f^{LR} = \mathbf{F}\left(\left[F_i^{LR}, \mathcal{P}_D\left(\gHR\right)\right]\right). \label{equ:recon}
\end{equation}

\Skip{
\subsubsection{\TBD{Dual shuffling -- fusion and} reconstruction network}\label{sec:reconst} We now describe the reconstruction network, which further process and extract features before the final upsampling. The current feature $F_\mathrm{current}$ is first extended from 32 channels to 64 channels by a single convolutional layer. Then, the extended current feature, historical features and HR pre-computed BRDF value $F_\beta$ are concatenated and fed into the reconstruction network:
\begin{align}
    F_\mathrm{extend} &= \textbf{Conv}_{32\to64}(F_\mathrm{current}) \\
    F_\mathrm{recon}^{LR} &= R\left(\left[F_\mathrm{extend}^{LR}, F_\mathrm{history}^{LR}, \mathcal{P}_I\left(F_\beta^{HR}\right)\right]\right) \label{equ:recon}
\end{align}
Noting that $F_\beta$ is an HR feature while $F_\mathrm{extend}$ and $F_\mathrm{history}$ are LR features, we explicitly use ``HR'' and ``LR'' superscripts accordingly in \cref{equ:recon} and subsequent equations for easier understandings. 
\TBD{As narrated in \cref{sec:shuf}, we leverage dual-shuffling technique to compactly fuse HR information into our LR reconstruction network and benefit from LR performance with negligible HR information loss. As the first half of dual shuffling (\ie fusion),}
$F_\beta$ is preprocessed by an inverted subpixel shuffling operation, namely ``$\mathrm{unshuffle}$'' in \cref{equ:recon} and visualized in \cref{fig:dual_shuffling}.

The reconstruction network consists of 6 convolution residual blocks preceded by an additional convolutional layer to adjust the input feature channels. As shown in \cref{fig:architectire}, a residual block consists of 2 convolutional layers whose input and output are summed. The number of hidden layer channels in all residual blocks within $R$ is unified to 128.
}



\paragraph{Final upscaling module.}
In this module, we skip-connect the first layer output $F_s^{LR}$ of the encoder $\mathbf{E}$ with the fused feature map $F_f^{LR}$ and then upscale this concatenated feature map via the pixel shuffling operation $\mathcal{P}_U$.
Finally, the upscaled feature map is processed by a single-layer convolutional network to obtain the RGB prediction of demodulated irradiance $\hat{L}_D^{HR}$ (\cref{sec:demod}):
\begin{align}
    \hat{L}_D^{HR} &= \mathbf{Conv}\left(\mathcal{P}_U\left(\left[F_s^{LR}, F_f^{LR}\right]\right)\right), \label{equ:shuffle}
\end{align}

After obtaining $\hat{L}_D^{HR}$, we can compute the final color prediction $\piHR$ by remodulating with $F_\beta^{HR}$ as described in \cref{equ:demod}.


\Skip{
\subsubsection{\TBD{Dual shuffling -- upsampling} and final layer.}
In this final stage, we concatenate the output of reconstruction network $F_\mathrm{recon}$ with primary feature $F_\mathrm{primary}$ as a skip connection. \TBD{As the second half of dual shuffling (\ie upsampling),} we leverage subpixel convolution proposed by ESPCN~\cite{shi2016real} to upsample the concatenated LR deep feature into a HR shallow feature $F_\mathrm{final}$ (\cref{equ:shuffle}). The HR shallow feature is then fed into a final single-layer convolutional network to obtain the final RGB prediction of demodulated irradiance $\hat{L}_D$ (\cref{sec:demod}). $\hat{L}_D$ is further multiplied by the pre-integrated BRDF $F_\beta$ to synthesize the final predicted HR image:
\begin{align}
    F_\mathrm{high}^{HR} &= U([F_\mathrm{recon}^{LR}, F_\mathrm{primary}^{LR}]), \label{equ:shuffle}\\
    \hat{L}_D^{HR} &= \textbf{Conv}_\mathrm{final}(F_\mathrm{final}^{HR}), \\
    \hat{I}^{HR} &= \hat{L}_D^{HR} \odot F_\beta^{HR}.
\end{align}
}


\Skip{
\subsubsection{History Reprojection Module}
Inspired by existing works~\cite{xiao2020neural,dlss,fsr,xess}, we also reuse historical frames to provide temporal information for the network so as to further enhance the prediction quality. The historical frames encoded by the exact same encoder (with shared weights) as in \cref{equ:encode} and the encoded features are reprojected onto current screen space according to the motion vectors.

\zjs{moved here from sec 3.4}
Nevertheless, the reprojected historical frames are not reliable everywhere due to the changes of dynamic scene and occlusion relationships, which can be visualized as ghosting and shimmering. In order to eliminate these artifacts, we design an attention-based network to localize and mask out unreliable regions in historical frames. We adopt the design of additive attention~\cite{bahdanau2014neural}:
\begin{equation}
    \label{equ:attention}
    \begin{split}
        H &= \mathrm{concatenate}(M,D_\mathrm{current},D_\mathrm{history}) \\
        A &= \sigma(\textbf{Conv}_\mathrm{v}(\tanh(\textbf{Conv}_\mathrm{a}(H)+\textbf{Conv}_\mathrm{b}(H))))
    \end{split}
\end{equation}
where $\sigma$ is the sigmoid activation function, $M$ is the motion vector, and $D_\mathrm{current},D_\mathrm{history}$ are depth maps of current and historical frames, respectively. The attention network generates an attention map $A$ that emphasizes regions of interest and masks out unreliable regions. $A$ will be further pixel-wise multiplied by the warped historical feature to obtain the masked historical feature $F_\mathrm{masked}$:
\begin{equation}\label{equ:mask}
    F_\mathrm{masked} = A \odot \mathrm{warp}(\textbf{Encode}(I_\mathrm{history}), M)
\end{equation}

\zjs{moved here from sec 3.5}
It is worth mentioning that we use the exact same encoder design with shared weights to process historical data, \ie the $\textbf{Encode}$ operation in \cref{equ:mask} and \cref{equ:encode} are exactly the same. In most previous works, \eg NSRR~\cite{xiao2020neural}, historical frames are encoded with different encoders from current frames, requiring additional convolutional computations and brings inevitable overheads in real-time rendering. Instead, our sharing encoder design makes it possible to cache the encoded features of previous frames and directly warp and reuse them in the current inference pass. This saves the encoding time of historical features, bringing significant performance optimizations.
}

\Skip{\zjs{To be confirmed}
We use the exact same encoder design with shared weights to process and reuse historical data, \ie the feature reuse mechanism described in \cref{sec:reuse_mechanism}. 
The historical features are fed into the attention network along with motion vectors (\cref{sec:attention,equ:attention}), and the generated attention map is multiplied with the historical feature to produce the masked feature $F_\mathrm{masked}$ as final output (\cref{equ:mask}), where unreliable occluded areas in historical frames are filtered out. Finally, the historical feature will be warped onto current screen space according to the motion vector $M$:
\begin{equation}
    F_\mathrm{history} = \mathrm{warp}(F_\mathrm{masked}, M)
\end{equation}}

\subsection{Training}
\label{sec:train}

We train our network with the supervision of ground truth high-resolution images, and the training process is end-to-end.

\paragraph{$L_1$ loss} We directly apply $L_1$ loss between HR ground truth $I^{HR}$ and the predicted HR image $\piHR$ to supervise the training:
\begin{equation}
    \mathcal{L}_\mathrm{color} = \| \piHR - I^{HR} \|_1.
\end{equation}

\paragraph{Perceptual loss} Introduced by \citet{johnson2016perceptual}, we use a pretrained VGG-16 network to compute perceptual loss to enhance semantic consistency:
\begin{equation}
    \mathcal{L}_\mathrm{perceptual} = \sum_{i\in V}{\| \mathrm{VGG}_i(\piHR) - \mathrm{VGG}_i(I^{HR}) \|_2},
\end{equation}
where $\mathrm{VGG}_i$ represents the i-th layer of pre-trained VGG network, and $V$ denotes the set of selected layers.

\paragraph{SSIM loss} Following NSRR~\cite{xiao2020neural}, we also incorporate structural similarity index (SSIM)~\cite{wang2004image} as an additional loss function: 
\begin{equation}
    \mathcal{L}_\mathrm{strutural} = 1 - \mathrm{SSIM}(\piHR, I^{HR}),
\end{equation}

\paragraph{Total loss} To summarize, the total loss is the weighted combination of all losses mentioned above:
\begin{equation}
    \mathcal{L} = \mathcal{L}_\mathrm{color} + \lambda_\mathrm{p}\mathcal{L}_\mathrm{perceptual} + \lambda_\mathrm{s}\mathcal{L}_\mathrm{strutural}.
\end{equation}
We empirically set $\lambda_\mathrm{p}=0.5$ and $\lambda_\mathrm{s}=0.05$ in our experiments.


\section{Experiment}

\begin{table*}[htbp]
\centering
\caption{Left: Quality comparisons with baselines in $4\times 4$ upsampling. The top 3 best results are highlighted with gold, silver, and bronze. Right: Quality comparisons in the challenging $8\times 8$ upsampling.
}
\begin{tabular}{ccccccccc||ccc}\hline
     & & Ours & Ours \Lightning & NSRR & MNSS & LIIF & FSR & XeSS & Ours-8x & NSRR-8x & MNSS-8x \\ \hline
     \multirow{4}{*}{\rotatebox{90}{PSNR (dB)}}
      & \texttt{Kite} & \cellcolor{gold}{32.33} & \cellcolor{silver}{31.22} & 27.74 & {28.00} & 26.47 & \cellcolor{bronze}{29.12} & 28.30 & \cellcolor{gold}{30.21} & 25.00 & 25.72 \\
      & \texttt{Showdown} & \cellcolor{gold}{36.32} & \cellcolor{silver}{31.42} & {30.27} & 29.17 & \cellcolor{bronze}{30.33} & 26.29 & 29.31 & \cellcolor{gold}{33.61} & 29.17 & 25.62 \\ 
      & \texttt{Slay} & \cellcolor{gold}{37.02} & {34.41} & \cellcolor{silver}{35.42} & \cellcolor{bronze}{35.39} & 31.12 & 32.39 & 34.94 & \cellcolor{gold}{34.26} & 32.12 & 33.47 \\ 
      & \texttt{City} & \cellcolor{gold}{28.94} & \cellcolor{silver}{28.66} & 27.65 & \cellcolor{bronze}{28.23} & 26.56 & 26.63 & 27.15 & \cellcolor{gold}{27.20} & 25.95 & 26.46 \\ \hline
     \multirow{4}{*}{\rotatebox{90}{SSIM}}
      & \texttt{Kite} & \cellcolor{gold}{0.933} & \cellcolor{silver}{0.900} & {0.832} & 0.829 & 0.817 & {0.887} & \cellcolor{bronze}{0.893} & \cellcolor{gold}{0.899} & 0.765 & 0.770 \\
      & \texttt{Showdown} & \cellcolor{gold}{0.976} & \cellcolor{silver}{0.949} & \cellcolor{bronze}{0.945} & 0.914 & 0.942 & 0.866 & 0.917 & \cellcolor{gold}{0.955} & 0.914 & 0.813 \\ 
      & \texttt{Slay} & \cellcolor{gold}{0.972} & {0.958} & \cellcolor{bronze}{0.962} & \cellcolor{silver}{0.963} & \cellcolor{bronze}{0.962} & 0.928 & 0.944 & \cellcolor{gold}{0.957} & 0.939 & 0.943 \\ 
      & \texttt{City} & \cellcolor{gold}{0.921} & \cellcolor{silver}{0.901} & \cellcolor{bronze}{0.899} & 0.896 & 0.874 & 0.836 & 0.888 & \cellcolor{gold}{0.916} & 0.873 & 0.873 \\ \hline
\end{tabular}
\label{tab:quality}
\end{table*}

In this section, we will demonstrate the comprehensive experimental results of our method.
We first introduce implementation details of our network and datasets used in training and testing (\cref{sec:detail}). Then, we compare our method with state-of-the-art methods in quality and performance (\cref{sec:comp}). \new{We further make a discussion on the performance and quality of integrating our method into rendering pipelines compared to native HR rendering in \cref{sec:native}.} At last, experiments are conducted to validate the effectiveness of each design (\cref{sec:ablation}).

\subsection{Implementation Details and Datasets}
\label{sec:detail}

\paragraph{Implementation details.} Our model is implemented and trained using PyTorch~\cite{paszke2019pytorch}. All training and testing are performed on a single NVIDIA RTX 3090 GPU.  For the best trade-off between performance and quality, we evaluate two implementations of our FuseSR model, namely \emph{Ours} and \emph{Ours \Lightning}\Skip{, in the following figures and tables}.
``Ours'' version provides full network implementation with optimal quality. 
``Our \Lightning'' version is more performance-oriented. It cuts off the history reusing module  in the encoder (\cref{sec:arch}), reduces the number of hidden channels of all layers in fusion network $\textbf{F}$ by half, and replaces all convolutional layers in $\textbf{F}$ with faster DWS layers~\cite{sandler2018mobilenetv2} to improve the performance at a slight sacrifice of quality.

\paragraph{Datasets.} We have constructed a large-scale dataset using four virtual scenes selected from Unreal Engine marketplace\footnote{\url{https://unrealengine.com/marketplace}}
\new{We select 2 scenes from Unreal Engine 4 (\texttt{Kite} and \texttt{Showdown}), and 2 scenes from Unreal Engine 5 (\texttt{Slay} and \texttt{City}).}
These scenes contain extensively complex geometry and lighting conditions with dynamic objects. Each scene contains 1080 continuous frames, of which 960 frames are used for training and 120 are used for testing. \new{To show our method's ability for high-quality rendering, the 2 scenes from UE 5 are rendered by real-time raytracing to achieve photorealistic appearances.}
We choose 4K ($3840\times2160$) as the resolution of the target HR frame and the LR frames are downscaled according to scale factor.
All frames are generated using UE 4~\shortcite{ue} and 5~\shortcite{ue5} with customized shaders to compute the required G-buffers, pre-integrated BRDF, and motion vectors.

\paragraph{Metrics.} Our evaluation comprises two aspects: performance and quality. We report the runtime of networks in milliseconds (ms) as a straightforward metric in performance. In terms of quality, we report two widely-used image metrics: peak signal-to-noise ratio (PSNR) and structural similarity index (SSIM)~\cite{wang2004image}. For both PSNR and SSIM, higher is better.

\subsection{Results and Comparisons}
\label{sec:comp}

\subsubsection{Quality evaluation.} 
We compare our method with several state-of-the-art super-resolution methods in both academia and industry, including \new{SOTA single image super-resolution method LIIF~\cite{chen2021learning}}, real-time rendering super-resolution methods NSRR~\cite{xiao2020neural} \new{and MNSS~\cite{MNSS}}, and methods widely used in the gaming industry including AMD’s FidelityFX\textsuperscript{\texttrademark} Super Resolution (FSR)~\cite{fsr} and Intel Xe Super Sampling (XeSS)~\cite{xess}.

In \cref{tab:quality}, we compare PSNR and SSIM metrics averaged over all frames from 4 scenes of our dataset \Skip{described in \cref{sec:detail}}.
Our method significantly outperforms other baselines in all scenes quantitatively. We also provide qualitative comparisons in \cref{fig:result_4x}. 
Our method faithfully produces HR details such as sharp edges and complex textures, owing to our exploitation of BRDF demodulation and effective H-Net design.


\subsubsection{Runtime performance.} The performance of a super-resolution method is crucial for its application to real-time high-resolution rendering.
After training, the optimized models are accelerated by NVIDIA TensorRT~\shortcite{tensorrt} with 16-bit mixed precision for optimal inference speed. 
In \cref{tab:runtime}, we report the total runtime of our method and baselines at different resolutions. 
Benefiting from our carefully designed H-Net architecture, both of our two versions achieve superior performance to NSRR at all configurations. 

\new{MNSS achieves competitive runtime performance due to its lightweight network design, especially at low resolutions. However, it requires network computations at HR level, leading to a more steep increase in running time compared to ours. With G-buffer generation time included, Ours \Lightning surpasses MNSS above 2K resolution.}

\new{The performance of our method is highly dependent on the upscaling factor, because our fusion network runs at LR level thanks to our pixel unshuffling alignment strategy. ``Ours-8x'' achieves superior performance to ``Ours'', especially for high resolutions.}

\begin{figure}[htbp]
    \centering
    \includegraphics[width=0.7\linewidth]{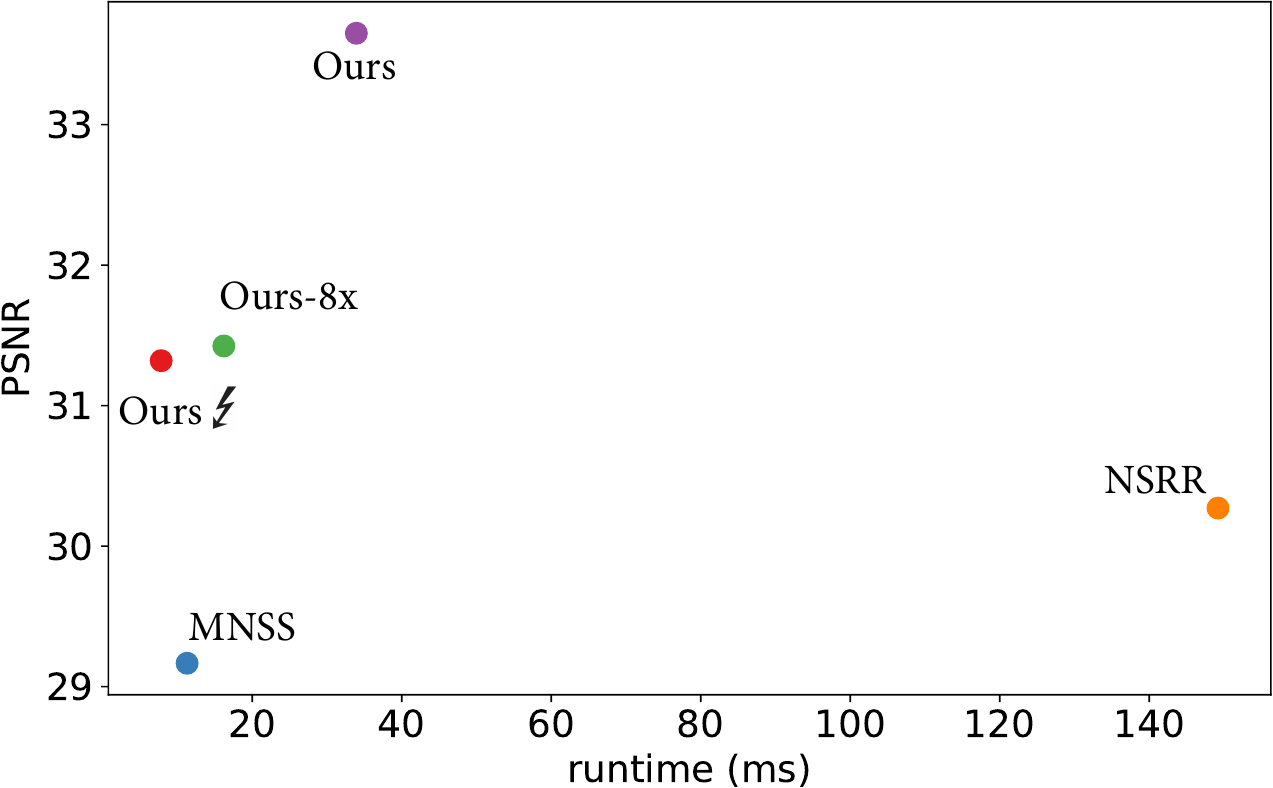}
    \caption{Comparisons on the trade-off between performance and quality.}
    \label{fig:pareto}
\end{figure}

``Ours \Lightning'' version indeed leads to a slight decrease in quality, which is reasonable due to the reduced network capacity. 
Nevertheless, it still performs on par with or better than other baselines. \new{\cref{fig:pareto} displays the trade-off between performance and quality.}

\begin{table}[htbp]
    \centering
    \caption{Comparison of total runtime and HR G-buffer generation time (in milliseconds) for SR tasks \new{($8\times8$ for ``Ours-8x'' and $4\times4$ for others)} with different target resolutions. Results are tested on an NVIDIA RTX 3090 GPU.}
    \begin{tabular}{ccccc}\hline
     & 720p & 1080p & 2K & 4K \\ \hline
    HR G-buffer & 0.83 & 0.93 & 1.97 & 2.35\\ \hline
    Ours & 6.21 & 8.44 & 15.09 & 33.93 \\
    Ours-8x & 6.20 & 7.57 & 8.79 & 16.20 \\
    Ours \Lightning & 2.66 & 2.88 & 3.96 & 7.82 \\
    NSRR & 13.53 & 26.29 & 64.02 & 149.20 \\
    MNSS & 2.26 & 3.57 & 5.52 & 11.29 \\ \hline
    \end{tabular}
    \label{tab:runtime}
\end{table}

{
\subsubsection{Upsampling Factor}
\new{Besides $4\times4$ super-resolution, we also evaluate our method on a more challenging $8\times8$ task, which requires the network to predict 64 HR sub-pixel colors from a single LR pixel. Such a high upsampling rate makes it nearly impossible for previous methods with the absence of  HR cues to recover any details, while our method still predicts high-fidelity results thanks to the demodulation and multi-resolution fusion. \cref{tab:quality} (Right) and \cref{fig:result_8x}  demonstrates quantitative and qualitative results of our method and SOTA baselines.}
}

\Skip{
\begin{table}[htbp]
    \centering
    \caption{Runtime Performance\zzh{To be refine. Deviation in total come from testing module separated overhead. NSRR takes 26.29ms in total in same device. HR Gbuffer takes 1.75ms according to Extra-Net}}
    \begin{tabular}{ccc}\hline
        Module & Ours (ms) & Ours \Lightning (ms)\\ \hline
        Encoder & 1.02 & 0.75 \\ \hline
        Warping & 0.14 & - \\ \hline
        Attention & 0.35 & - \\ \hline
        Reconstruction & 5.57 & 1.15 \\ \hline
        Sub-pixel Upsampling &  \multicolumn{2}{c}{1.32} \\ \hline
        Total & 8.44 & 2.88 \\ \hline
    \end{tabular}
\end{table}
}


{
\subsection{Discussion on Real-time Rendering}
\label{sec:native}

\paragraph{G-buffer generation.} We test G-buffer generation time by recording the rendering state via the UE built-in profiling tool\Skip{ GPU visualizer and ``Insights''}. For each scene, we measure the average G-buffer time on a sequence of dynamic frames and take 10 samples to cover as many scenarios as possible. Notably, G-buffer generation time is affected by various factors and there is a variance between certain frames within four orders of magnitude ranging from 0.01 ms to less than 10 ms. In general, the average G-buffer time is significantly lower than network time (as reported in the first row of \cref{tab:runtime}). In modern game engines, simplification techniques, such as LOD or new techniques like Nanite in UE5, can further reduce the overhead of G-Buffer generation to a stably low cost.

\begin{figure}[htbp]
    \centering
    \includegraphics[width=\linewidth]{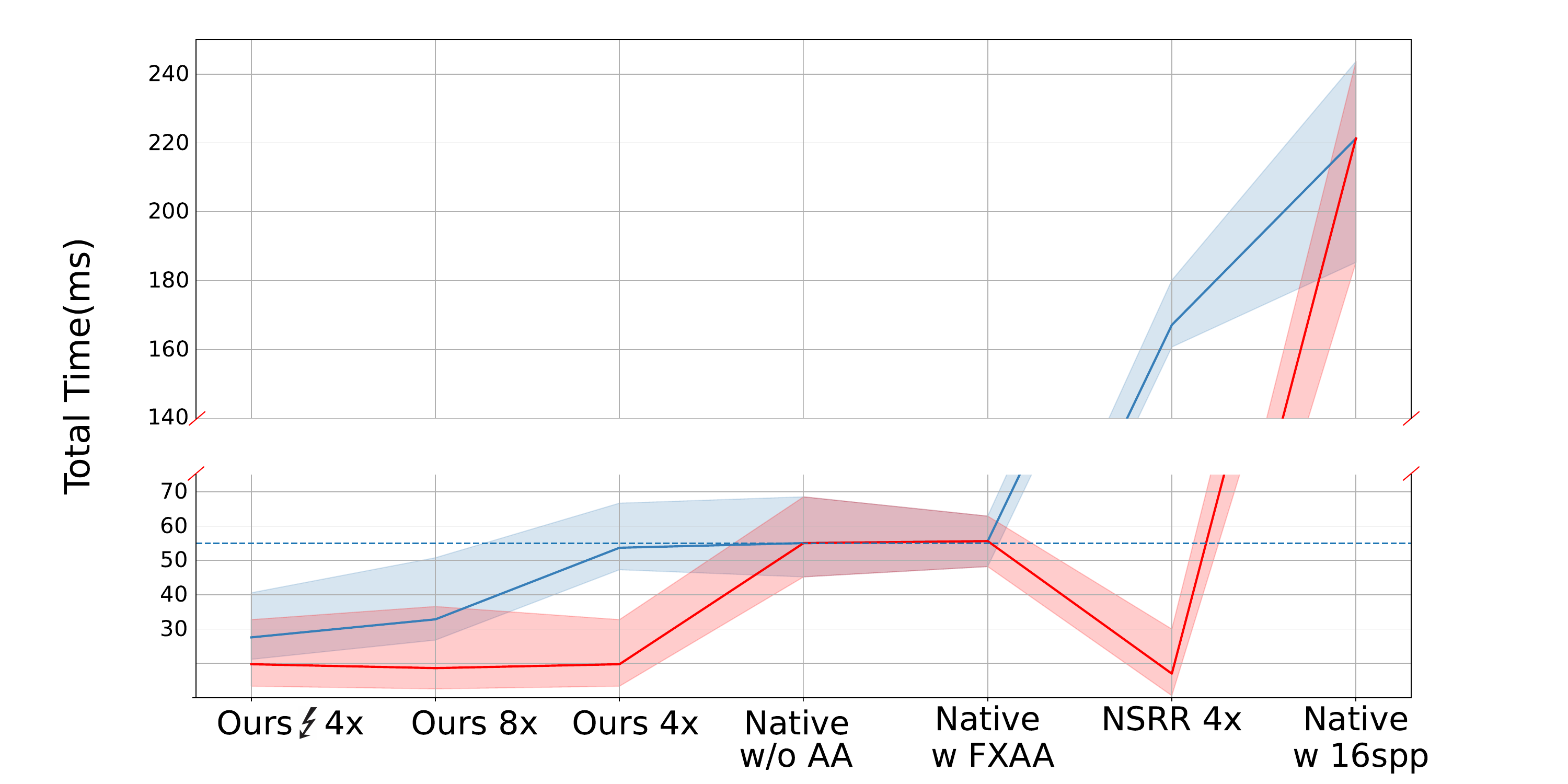}
    \caption{Comparison of runtime between our method and native rendering of HR image at 4K resolution. Blue lines denote the total runtime (LR rendering + G-buffer generation + network inference), while red lines denote the overhead within the rendering pipeline (LR rendering + G-buffer). Results are taken from \texttt{Slay} scene in UE5.}
    \label{fig:native}
\end{figure}

\paragraph{Comparisons to HR native rendering.} The ultimate goal of SR for real-time rendering is to reduce computational overhead by avoiding native rendering of HR images by the rendering engine. The performance of native rendering is affected by multiple factors including resolution and anti-aliasing options. 
In \cref{fig:native}, we compare the runtime of HR native rendering with the total runtime of rendering an LR image and then super-resolving it into HR by our method. We report native rendering runtimes with different anti-aliasing levels (w/o AA, FXAA, and SSAA), and our results with different network configurations (Ours and Ours \Lightning) and different SR upscaling factors (4x and 8x). 
Results show that integrating our method into the rendering pipeline indeed reduces computational overhead compared to native rendering.

\paragraph{Scene complexity} The native rendering overhead typically depends on the complexity of scenes and lighting conditions, while the inference time of SR network basically remains constant among varying scenes. Integrating SR into rendering pipelines can increase the potential of designing more complex scenes for modern games without noticeable degradation of rendering performance.


}

\Skip{
\subsection{Generalizability and Upsampling Factor}
\label{sec:GUT}
\subsubsection{Scene Generalizability} Due to the inherent scene diversity, the SR quality varies among different scenes. For example, \texttt{ZenGarden} mainly consists of large diffuse surfaces, making it least challenging for reconstruction, while \texttt{Kite} is an outdoor scene with bright sunlights and complex grasses and leaves, making it most challenging. We choose to train per scene to maximize its quality, while our method also generalizes well across scenes. In \cref{tab:quality} (right), we report the PSNR and SSIM comparisons in scene generalizability. 

The quantitative results of ``Ours-All'' and ``Ours-AllButOne'' in \cref{tab:quality} indicate that our method has good generalizability across different scenes.
Compared to the per-scene version, the results only decrease by a small margin, showing that the network can learn useful features from other scenes to enhance quality. 

In addition, we also compare our generalizability with NSRR, as shown in the ``NSRR-All'' column of \cref{tab:quality}, and it turns out that our method is superior.

\subsubsection{Upsampling Factor} Albeit we mainly narrate our method under a $4\times4$ SR task, we also test on other upsampling factors, specifically, $2\times2$ and $8\times8$. We keep the target resolution at 4K and adjust the input resolution accordingly. The quantitative results are provided in \cref{tab:ratio}.

\begin{table}[htbp]
    \centering
    \caption{Quantitative comparisons over SR with different upsampling factors on \texttt{ZenGarden} scene.}
    \begin{tabular}{ccccccc}\hline
        \multirow{2}{*}{Ratio} & \multicolumn{2}{c}{Ours} & \multicolumn{2}{c}{NSRR} & \multicolumn{2}{c}{RCAN} \\
         & PSNR & SSIM & PSNR & SSIM & PSNR & SSIM \\ \hline
        $2\times2$ & \cellcolor{gold}50.64 & \cellcolor{gold}0.998 & 39.71 & 0.987 & 39.52 & 0.975 \\
        $4\times4$ & \cellcolor{gold}49.08 & \cellcolor{gold}0.997 & 37.67 & 0.974 & 38.04 & 0.974 \\
        $8\times8$ & \cellcolor{gold}45.29 & \cellcolor{gold}0.993 & 35.05 & 0.960 & 30.57 & 0.931 \\ \hline
    \end{tabular}
    \label{tab:ratio}
\end{table}

Reasonably, the reconstruction quality improves as the upsampling factor reduces. In the extremely challenging $8\times8$ SR task, NSRR fails to preserve sharp edges, thin objects, or textured areas, while our method still yields a high-quality reconstruction (see \cref{fig:ratio}). 

\begin{figure}[htbp]
    \centering
    {\setlength{\tabcolsep}{2pt}
    \begin{tabular}{cccc}
        \includegraphics[width=0.2\linewidth]{images/result/8x/Input_aoi0.png} & \includegraphics[width=0.2\linewidth]{images/result/8x/Ours_aoi0.png} & \includegraphics[width=0.2\linewidth]{images/result/8x/NSRR_aoi0.png} & \includegraphics[width=0.2\linewidth]{images/result/8x/Reference_aoi0.png} \\
        \includegraphics[width=0.2\linewidth]{images/result/8x/Input_aoi2.png} & \includegraphics[width=0.2\linewidth]{images/result/8x/Ours_aoi2.png} & \includegraphics[width=0.2\linewidth]{images/result/8x/NSRR_aoi2.png} & \includegraphics[width=0.2\linewidth]{images/result/8x/Reference_aoi2.png}\\
        Input & Ours & NSRR & Reference \\
    \end{tabular}
    }
    \caption{Qualitative comparisons of $8\times8$ SR on \texttt{ZenGarden} scene. Complete images are presented in the supplement.}
    \label{fig:ratio}
\end{figure}
}

\subsection{Ablation Studies}
\label{sec:ablation}

\subsubsection{Network modules}\label{sec:ab_module}
In \cref{tab:alblation}, we report ablation experiments on our BRDF demodulation (\cref{sec:demod}) and HR G-buffer fusion (\cref{sec:shuf}).
The results confirm the effectiveness of our design. 

\begin{table}[htbp]
    \centering
    \caption{Ablation studies for HR G-buffer fusion and BRDF demodulation. Results in the table are averaged from experiments on \texttt{Kite} and \texttt{Slay} scenes.}
    \begin{tabular}{cccc}\hline
        HR G-Buffer & Demodulation & PSNR(dB) & SSIM \\ \hline
        \xmark & \xmark & 31.88 & 0.891 \\ \hline
        \cmark & \xmark & 33.74 & 0.941 \\ \hline
        \xmark & \cmark & 33.05 & 0.926 \\ \hline
        \cmark & \cmark & 34.67 & 0.952 \\ \hline
    \end{tabular}
    \label{tab:alblation}
\end{table}

{
\subsubsection{Alignment strategy}\label{sec:ab_align}
\new{\cref{tab:ab_align,fig:ab_align} shows the comparisons between different alignment strategies used before the fusion network. Our strategy significantly outperforms maximum pooling and average pooling due to the preservation of HR details. Our strategy even outperforms the upsampling alignment strategy (where the fusion network runs much slower at HR level), showing the effectiveness of compact LR implicit representations.}
}

\begin{table}[htbp]
    \centering
    \caption{Ablation studies for different alignment strategies on 2 scenes.}
    \begin{tabular}{ccccc}\hline
         & Ours & Avg-Pool & Max-Pool & Upsampling \\\hline
        \texttt{Kite} & 32.33 & 32.14 & 31.87 & 32.09 \\\hline
        \texttt{Slay} & 37.02 & 36.32 & 36.11 & 36.69 \\\hline
    \end{tabular}
    \label{tab:ab_align}
\end{table}

{
\setlength{\tabcolsep}{2pt}
\begin{figure}[h!]
    \centering
    \begin{tabular}{ccc}
        \stackinset{l}{1pt}{t}{1pt}{\textcolor{white}{Ours}}{\includegraphics[width=0.28\linewidth]{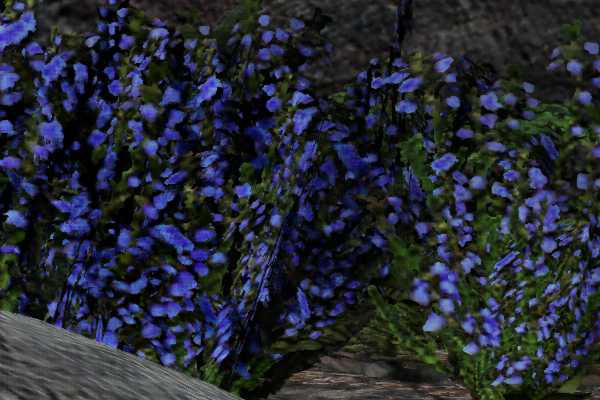}} & \stackinset{l}{1pt}{t}{1pt}{\textcolor{white}{Upsampling}}{\includegraphics[width=0.28\linewidth]{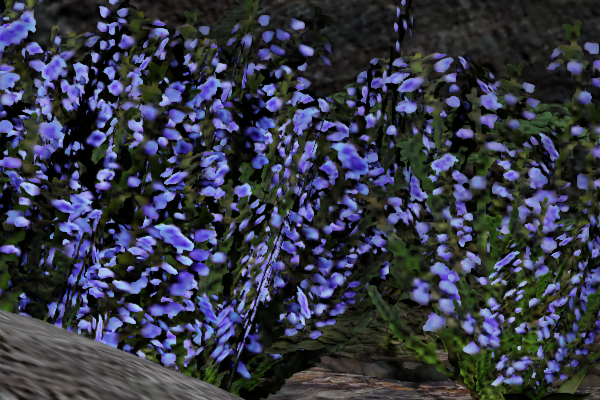}} & \stackinset{l}{1pt}{t}{1pt}{\textcolor{white}{Reference}}{\includegraphics[width=0.28\linewidth]{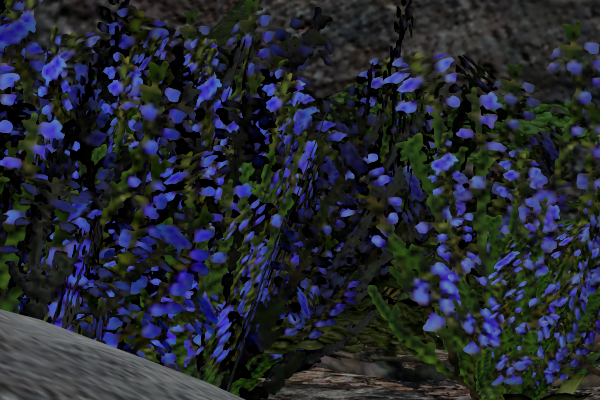}} \\
        \stackinset{l}{1pt}{t}{1pt}{\textcolor{white}{Ours}}{\includegraphics[width=0.28\linewidth]{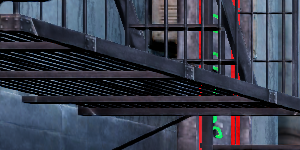}} & \stackinset{l}{1pt}{t}{1pt}{\textcolor{white}{Avg-Pool}}{\includegraphics[width=0.28\linewidth]{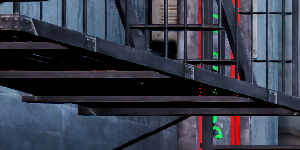}} & \stackinset{l}{1pt}{t}{1pt}{\textcolor{white}{Reference}}{\includegraphics[width=0.28\linewidth]{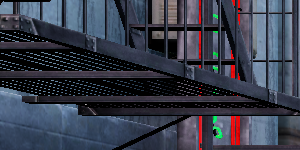}}
    \end{tabular}
    \caption{Ablation studies for alignment strategies. Upsampling leads to color bias, while spatial pooling leads to the loss of thin structures.}
    \label{fig:ab_align}
\end{figure}
}

\section{Discussion and Conclusion}
\label{sec:limit}

{
\paragraph{Integration cost.} While HR G-buffers provide a good guidance for upscaling and the demodulation further helps with the quality, these additional inputs require major modifications to modern game engines to integrate FuseSR. For example, to provide pre-integrated BRDF term $F_\beta(\bwo)$, the engine should generate pre-integrated BRDF LUT for each material asset. To reduce integration costs, some rational trade-offs may be considered such as using FuseSR without demodulation or replacing $F_\beta(\bwo)$ with albedo.

}

\paragraph{Material Generalization.}
Split-sum approximation offers an efficient means for BRDF demodulation, but it also limits the material types we can support.
Materials that violate the assumptions of split-sum approximation, such as translucent and anisotropic materials, are not available in our method.
Therefore, finding a more general BRDF demodulation method is an important direction for our future research.


\paragraph{Potential of H-Net.} 
We believe that our H-Net design \Skip{and dual-shuffling technique} are flexible and scalable, with the potential to be extended to more resolution levels and more multi-resolution tasks such as multi-scale object detection. We leave it as an interesting future work.

\paragraph{Conclusion}
We present FuseSR, an efficient and effective super-resolution network that predicts high-quality $4 \times 4$ (even $8 \times 8$) upsampled reconstructions according to the corresponding low-resolution frames.
With introducing pre-integrated BRDF demodulation and H-Net design, our method achieves a new SOTA for real-time superresolution.
The experiments show that our method not only strikes an outstanding balance strikes an excellent balance between quality and performance, but also exhibits excellent generalizability and temporal stability. We hope our work will stimulate further developments of neural superresolution.

\begin{acks}
The work was partially supported by Key R\&D Program of Zhejiang Province (No. 2023C01039), Zhejiang Lab (121005-PI2101), and Information Technology Center and State Key Lab of CAD\&CG, Zhejiang University.
\end{acks}

\newpage

\bibliographystyle{ACM-Reference-Format}
\bibliography{ref}


\begin{thebibliography}{29}


\ifx \showCODEN    \undefined \def \showCODEN     #1{\unskip}     \fi
\ifx \showDOI      \undefined \def \showDOI       #1{#1}\fi
\ifx \showISBNx    \undefined \def \showISBNx     #1{\unskip}     \fi
\ifx \showISBNxiii \undefined \def \showISBNxiii  #1{\unskip}     \fi
\ifx \showISSN     \undefined \def \showISSN      #1{\unskip}     \fi
\ifx \showLCCN     \undefined \def \showLCCN      #1{\unskip}     \fi
\ifx \shownote     \undefined \def \shownote      #1{#1}          \fi
\ifx \showarticletitle \undefined \def \showarticletitle #1{#1}   \fi
\ifx \showURL      \undefined \def \showURL       {\relax}        \fi
\providecommand\bibfield[2]{#2}
\providecommand\bibinfo[2]{#2}
\providecommand\natexlab[1]{#1}
\providecommand\showeprint[2][]{arXiv:#2}

\bibitem[Akeley(1993)]%
        {akeley1993reality}
\bibfield{author}{\bibinfo{person}{Kurt Akeley}.}
  \bibinfo{year}{1993}\natexlab{}.
\newblock \showarticletitle{Reality engine graphics}. In
  \bibinfo{booktitle}{\emph{Proceedings of the 20th annual conference on
  Computer graphics and interactive techniques}}. \bibinfo{pages}{109--116}.
\newblock


\bibitem[{AMD}(2021)]%
        {fsr}
\bibfield{author}{\bibinfo{person}{{AMD}}.} \bibinfo{year}{2021}\natexlab{}.
\newblock \bibinfo{booktitle}{\emph{AMD
  FidelityFX\textsuperscript{\texttrademark} Super Resolution}}.
\newblock
\urldef\tempurl%
\url{https://www.amd.com/en/technologies/fidelityfx-super-resolution/}
\showURL{%
\tempurl}


\bibitem[Bako et~al\mbox{.}(2017)]%
        {bako2017kernel}
\bibfield{author}{\bibinfo{person}{Steve Bako}, \bibinfo{person}{Thijs Vogels},
  \bibinfo{person}{Brian McWilliams}, \bibinfo{person}{Mark Meyer},
  \bibinfo{person}{Jan Nov{\'a}k}, \bibinfo{person}{Alex Harvill},
  \bibinfo{person}{Pradeep Sen}, \bibinfo{person}{Tony Derose}, {and}
  \bibinfo{person}{Fabrice Rousselle}.} \bibinfo{year}{2017}\natexlab{}.
\newblock \showarticletitle{Kernel-predicting convolutional networks for
  denoising Monte Carlo renderings.}
\newblock \bibinfo{journal}{\emph{ACM Trans. Graph.}} \bibinfo{volume}{36},
  \bibinfo{number}{4} (\bibinfo{year}{2017}), \bibinfo{pages}{97--1}.
\newblock


\bibitem[Chen et~al\mbox{.}(2021)]%
        {chen2021learning}
\bibfield{author}{\bibinfo{person}{Yinbo Chen}, \bibinfo{person}{Sifei Liu},
  {and} \bibinfo{person}{Xiaolong Wang}.} \bibinfo{year}{2021}\natexlab{}.
\newblock \showarticletitle{Learning continuous image representation with local
  implicit image function}. In \bibinfo{booktitle}{\emph{Proceedings of the
  IEEE/CVF conference on computer vision and pattern recognition}}.
  \bibinfo{pages}{8628--8638}.
\newblock


\bibitem[{Epic Games}(2020a)]%
        {taau}
\bibfield{author}{\bibinfo{person}{{Epic Games}}.}
  \bibinfo{year}{2020}\natexlab{a}.
\newblock \bibinfo{booktitle}{\emph{Screen Percentage with Temporal Upscale in
  Unreal Engine}}.
\newblock
\urldef\tempurl%
\url{https://docs.unrealengine.com/en-US/screen-percentage-with-temporal-upscale-in-unreal-engine/}
\showURL{%
\tempurl}


\bibitem[{Epic Games}(2020b)]%
        {ue}
\bibfield{author}{\bibinfo{person}{{Epic Games}}.}
  \bibinfo{year}{2020}\natexlab{b}.
\newblock \bibinfo{booktitle}{\emph{Unreal Engine}}.
\newblock
\urldef\tempurl%
\url{https://www.unrealengine.com/}
\showURL{%
\tempurl}


\bibitem[{Epic Games}(2021)]%
        {ue5}
\bibfield{author}{\bibinfo{person}{{Epic Games}}.}
  \bibinfo{year}{2021}\natexlab{}.
\newblock \bibinfo{title}{Unreal Engine}.
\newblock
  \bibinfo{howpublished}{\url{https://www.unrealengine.com/en-US/unreal-engine-5}}.
\newblock


\bibitem[Fan et~al\mbox{.}(2021)]%
        {fan2021real}
\bibfield{author}{\bibinfo{person}{Hangming Fan}, \bibinfo{person}{Rui Wang},
  \bibinfo{person}{Yuchi Huo}, {and} \bibinfo{person}{Hujun Bao}.}
  \bibinfo{year}{2021}\natexlab{}.
\newblock \showarticletitle{Real-time Monte Carlo Denoising with Weight Sharing
  Kernel Prediction Network}. In \bibinfo{booktitle}{\emph{Computer Graphics
  Forum}}, Vol.~\bibinfo{volume}{40}. Wiley Online Library,
  \bibinfo{pages}{15--27}.
\newblock


\bibitem[Gharbi et~al\mbox{.}(2016)]%
        {gharbi2016deep}
\bibfield{author}{\bibinfo{person}{Micha{\"e}l Gharbi}, \bibinfo{person}{Gaurav
  Chaurasia}, \bibinfo{person}{Sylvain Paris}, {and} \bibinfo{person}{Fr{\'e}do
  Durand}.} \bibinfo{year}{2016}\natexlab{}.
\newblock \showarticletitle{Deep joint demosaicking and denoising}.
\newblock \bibinfo{journal}{\emph{ACM Transactions on Graphics (ToG)}}
  \bibinfo{volume}{35}, \bibinfo{number}{6} (\bibinfo{year}{2016}),
  \bibinfo{pages}{1--12}.
\newblock


\bibitem[Guo et~al\mbox{.}(2021)]%
        {guo2021extranet}
\bibfield{author}{\bibinfo{person}{Jie Guo}, \bibinfo{person}{Xihao Fu},
  \bibinfo{person}{Liqiang Lin}, \bibinfo{person}{Hengjun Ma},
  \bibinfo{person}{Yanwen Guo}, \bibinfo{person}{Shiqiu Liu}, {and}
  \bibinfo{person}{Ling-Qi Yan}.} \bibinfo{year}{2021}\natexlab{}.
\newblock \showarticletitle{ExtraNet: real-time extrapolated rendering for
  low-latency temporal supersampling}.
\newblock \bibinfo{journal}{\emph{ACM Transactions on Graphics (TOG)}}
  \bibinfo{volume}{40}, \bibinfo{number}{6} (\bibinfo{year}{2021}),
  \bibinfo{pages}{1--16}.
\newblock


\bibitem[{Intel}(2022)]%
        {xess}
\bibfield{author}{\bibinfo{person}{{Intel}}.} \bibinfo{year}{2022}\natexlab{}.
\newblock \bibinfo{booktitle}{\emph{Intel Xe Super Sampling}}.
\newblock
\urldef\tempurl%
\url{https://www.intel.com/content/www/us/en/products/docs/arc-discrete-graphics/xess.html/}
\showURL{%
\tempurl}


\bibitem[Johnson et~al\mbox{.}(2016)]%
        {johnson2016perceptual}
\bibfield{author}{\bibinfo{person}{Justin Johnson}, \bibinfo{person}{Alexandre
  Alahi}, {and} \bibinfo{person}{Li Fei-Fei}.} \bibinfo{year}{2016}\natexlab{}.
\newblock \showarticletitle{Perceptual losses for real-time style transfer and
  super-resolution}. In \bibinfo{booktitle}{\emph{European conference on
  computer vision}}. Springer, \bibinfo{pages}{694--711}.
\newblock


\bibitem[Kajiya(1986)]%
        {kajiya1986rendering}
\bibfield{author}{\bibinfo{person}{James~T Kajiya}.}
  \bibinfo{year}{1986}\natexlab{}.
\newblock \showarticletitle{The rendering equation}. In
  \bibinfo{booktitle}{\emph{Proceedings of the 13th annual conference on
  Computer graphics and interactive techniques}}. \bibinfo{pages}{143--150}.
\newblock


\bibitem[Kaplanyan et~al\mbox{.}(2019)]%
        {kaplanyan2019deepfovea}
\bibfield{author}{\bibinfo{person}{Anton~S Kaplanyan}, \bibinfo{person}{Anton
  Sochenov}, \bibinfo{person}{Thomas Leimk{\"u}hler}, \bibinfo{person}{Mikhail
  Okunev}, \bibinfo{person}{Todd Goodall}, {and} \bibinfo{person}{Gizem Rufo}.}
  \bibinfo{year}{2019}\natexlab{}.
\newblock \showarticletitle{DeepFovea: Neural reconstruction for foveated
  rendering and video compression using learned statistics of natural videos}.
\newblock \bibinfo{journal}{\emph{ACM Transactions on Graphics (TOG)}}
  \bibinfo{volume}{38}, \bibinfo{number}{6} (\bibinfo{year}{2019}),
  \bibinfo{pages}{1--13}.
\newblock


\bibitem[Karis(2013)]%
        {karis2013}
\bibfield{author}{\bibinfo{person}{Brian Karis}.}
  \bibinfo{year}{2013}\natexlab{}.
\newblock \showarticletitle{Real shading in unreal engine 4}. In
  \bibinfo{booktitle}{\emph{SIGGRAPH Courses: Physically Based Shading in
  Theory and Practice}}.
\newblock


\bibitem[Karis(2014)]%
        {karis2014}
\bibfield{author}{\bibinfo{person}{Brian Karis}.}
  \bibinfo{year}{2014}\natexlab{}.
\newblock \showarticletitle{High Quality Temporal Anti-Aliasing}. In
  \bibinfo{booktitle}{\emph{SIGGRAPH Courses: Advances in Real-Time
  Rendering}}.
\newblock


\bibitem[NVIDIA(2018)]%
        {dlss}
NVIDIA \bibinfo{year}{2018}\natexlab{}.
\newblock \bibinfo{booktitle}{\emph{Deep Learning Super Sampling (DLSS)
  Technology | NVIDIA}}.
\newblock NVIDIA.
\newblock
\urldef\tempurl%
\url{https://www.nvidia.com/en-us/geforce/technologies/dlss/}
\showURL{%
\tempurl}


\bibitem[{NVIDIA}(2018)]%
        {tensorrt}
\bibfield{author}{\bibinfo{person}{{NVIDIA}}.} \bibinfo{year}{2018}\natexlab{}.
\newblock \bibinfo{booktitle}{\emph{TensorRT}}.
\newblock
\urldef\tempurl%
\url{https://developer.nvidia.com/tensorrt/}
\showURL{%
\tempurl}


\bibitem[Paszke et~al\mbox{.}(2019)]%
        {paszke2019pytorch}
\bibfield{author}{\bibinfo{person}{Adam Paszke}, \bibinfo{person}{Sam Gross},
  \bibinfo{person}{Francisco Massa}, \bibinfo{person}{Adam Lerer},
  \bibinfo{person}{James Bradbury}, \bibinfo{person}{Gregory Chanan},
  \bibinfo{person}{Trevor Killeen}, \bibinfo{person}{Zeming Lin},
  \bibinfo{person}{Natalia Gimelshein}, \bibinfo{person}{Luca Antiga},
  {et~al\mbox{.}}} \bibinfo{year}{2019}\natexlab{}.
\newblock \showarticletitle{Pytorch: An imperative style, high-performance deep
  learning library}.
\newblock \bibinfo{journal}{\emph{Advances in neural information processing
  systems}}  \bibinfo{volume}{32} (\bibinfo{year}{2019}).
\newblock


\bibitem[Sandler et~al\mbox{.}(2018)]%
        {sandler2018mobilenetv2}
\bibfield{author}{\bibinfo{person}{Mark Sandler}, \bibinfo{person}{Andrew
  Howard}, \bibinfo{person}{Menglong Zhu}, \bibinfo{person}{Andrey Zhmoginov},
  {and} \bibinfo{person}{Liang-Chieh Chen}.} \bibinfo{year}{2018}\natexlab{}.
\newblock \showarticletitle{Mobilenetv2: Inverted residuals and linear
  bottlenecks}. In \bibinfo{booktitle}{\emph{Proceedings of the IEEE conference
  on computer vision and pattern recognition}}. \bibinfo{pages}{4510--4520}.
\newblock


\bibitem[Schied et~al\mbox{.}(2017)]%
        {schied2017spatiotemporal}
\bibfield{author}{\bibinfo{person}{Christoph Schied}, \bibinfo{person}{Anton
  Kaplanyan}, \bibinfo{person}{Chris Wyman}, \bibinfo{person}{Anjul Patney},
  \bibinfo{person}{Chakravarty R~Alla Chaitanya}, \bibinfo{person}{John
  Burgess}, \bibinfo{person}{Shiqiu Liu}, \bibinfo{person}{Carsten
  Dachsbacher}, \bibinfo{person}{Aaron Lefohn}, {and} \bibinfo{person}{Marco
  Salvi}.} \bibinfo{year}{2017}\natexlab{}.
\newblock \showarticletitle{Spatiotemporal variance-guided filtering: real-time
  reconstruction for path-traced global illumination}.
\newblock In \bibinfo{booktitle}{\emph{Proceedings of High Performance
  Graphics}}. \bibinfo{pages}{1--12}.
\newblock


\bibitem[Shi et~al\mbox{.}(2016)]%
        {shi2016real}
\bibfield{author}{\bibinfo{person}{Wenzhe Shi}, \bibinfo{person}{Jose
  Caballero}, \bibinfo{person}{Ferenc Husz{\'a}r}, \bibinfo{person}{Johannes
  Totz}, \bibinfo{person}{Andrew~P Aitken}, \bibinfo{person}{Rob Bishop},
  \bibinfo{person}{Daniel Rueckert}, {and} \bibinfo{person}{Zehan Wang}.}
  \bibinfo{year}{2016}\natexlab{}.
\newblock \showarticletitle{Real-time single image and video super-resolution
  using an efficient sub-pixel convolutional neural network}. In
  \bibinfo{booktitle}{\emph{Proceedings of the IEEE conference on computer
  vision and pattern recognition}}. \bibinfo{pages}{1874--1883}.
\newblock


\bibitem[Wang et~al\mbox{.}(2004)]%
        {wang2004image}
\bibfield{author}{\bibinfo{person}{Zhou Wang}, \bibinfo{person}{Alan~C Bovik},
  \bibinfo{person}{Hamid~R Sheikh}, {and} \bibinfo{person}{Eero~P Simoncelli}.}
  \bibinfo{year}{2004}\natexlab{}.
\newblock \showarticletitle{Image quality assessment: from error visibility to
  structural similarity}.
\newblock \bibinfo{journal}{\emph{IEEE transactions on image processing}}
  \bibinfo{volume}{13}, \bibinfo{number}{4} (\bibinfo{year}{2004}),
  \bibinfo{pages}{600--612}.
\newblock


\bibitem[Xiao et~al\mbox{.}(2020)]%
        {xiao2020neural}
\bibfield{author}{\bibinfo{person}{Lei Xiao}, \bibinfo{person}{Salah Nouri},
  \bibinfo{person}{Matt Chapman}, \bibinfo{person}{Alexander Fix},
  \bibinfo{person}{Douglas Lanman}, {and} \bibinfo{person}{Anton Kaplanyan}.}
  \bibinfo{year}{2020}\natexlab{}.
\newblock \showarticletitle{Neural supersampling for real-time rendering}.
\newblock \bibinfo{journal}{\emph{ACM Transactions on Graphics (TOG)}}
  \bibinfo{volume}{39}, \bibinfo{number}{4} (\bibinfo{year}{2020}),
  \bibinfo{pages}{142--1}.
\newblock


\bibitem[Yang et~al\mbox{.}(2020)]%
        {yang2020survey}
\bibfield{author}{\bibinfo{person}{Lei Yang}, \bibinfo{person}{Shiqiu Liu},
  {and} \bibinfo{person}{Marco Salvi}.} \bibinfo{year}{2020}\natexlab{}.
\newblock \showarticletitle{A survey of temporal antialiasing techniques}. In
  \bibinfo{booktitle}{\emph{Computer graphics forum}},
  Vol.~\bibinfo{volume}{39}. Wiley Online Library, \bibinfo{pages}{607--621}.
\newblock


\bibitem[Yang et~al\mbox{.}(2023)]%
        {MNSS}
\bibfield{author}{\bibinfo{person}{Sipeng Yang}, \bibinfo{person}{Yunlu Zhao},
  \bibinfo{person}{Yuzhe Luo}, \bibinfo{person}{He Wang},
  \bibinfo{person}{Hongyu Sun}, \bibinfo{person}{Chen Li},
  \bibinfo{person}{Binghuang Cai}, {and} \bibinfo{person}{Xiaogang Jin}.}
  \bibinfo{year}{2023}\natexlab{}.
\newblock \showarticletitle{MNSS: Neural Supersampling Framework for Real-Time
  Rendering on Mobile Devices}.
\newblock \bibinfo{journal}{\emph{IEEE Transactions on Visualization and
  Computer Graphics}} (\bibinfo{year}{2023}), \bibinfo{pages}{1--14}.
\newblock
\urldef\tempurl%
\url{https://doi.org/10.1109/TVCG.2023.3259141}
\showDOI{\tempurl}


\bibitem[Young(2006)]%
        {yong2006csaa}
\bibfield{author}{\bibinfo{person}{Peter Young}.}
  \bibinfo{year}{2006}\natexlab{}.
\newblock \bibinfo{booktitle}{\emph{Coverage sampled anti-aliasing}}.
\newblock \bibinfo{type}{{T}echnical {R}eport}. \bibinfo{institution}{NVIDIA
  Corporation}.
\newblock


\bibitem[Zeng et~al\mbox{.}(2021)]%
        {zeng2021temporally}
\bibfield{author}{\bibinfo{person}{Zheng Zeng}, \bibinfo{person}{Shiqiu Liu},
  \bibinfo{person}{Jinglei Yang}, \bibinfo{person}{Lu Wang}, {and}
  \bibinfo{person}{Ling-Qi Yan}.} \bibinfo{year}{2021}\natexlab{}.
\newblock \showarticletitle{Temporally Reliable Motion Vectors for Real-time
  Ray Tracing}. In \bibinfo{booktitle}{\emph{Computer Graphics Forum}},
  Vol.~\bibinfo{volume}{40}. Wiley Online Library, \bibinfo{pages}{79--90}.
\newblock


\bibitem[Zhuang et~al\mbox{.}(2021)]%
        {zhuang2021real}
\bibfield{author}{\bibinfo{person}{Tao Zhuang}, \bibinfo{person}{Pengfei Shen},
  \bibinfo{person}{Beibei Wang}, {and} \bibinfo{person}{Ligang Liu}.}
  \bibinfo{year}{2021}\natexlab{}.
\newblock \showarticletitle{Real-time Denoising Using BRDF Pre-integration
  Factorization}. In \bibinfo{booktitle}{\emph{Computer Graphics Forum}},
  Vol.~\bibinfo{volume}{40}. Wiley Online Library, \bibinfo{pages}{173--180}.
\newblock


\end{thebibliography}


\begin{figure*}[htbp]
    \setlength{\tabcolsep}{2pt}
    \centering
    \begin{tabular}{cc}
        \includegraphics[width=0.49\linewidth]{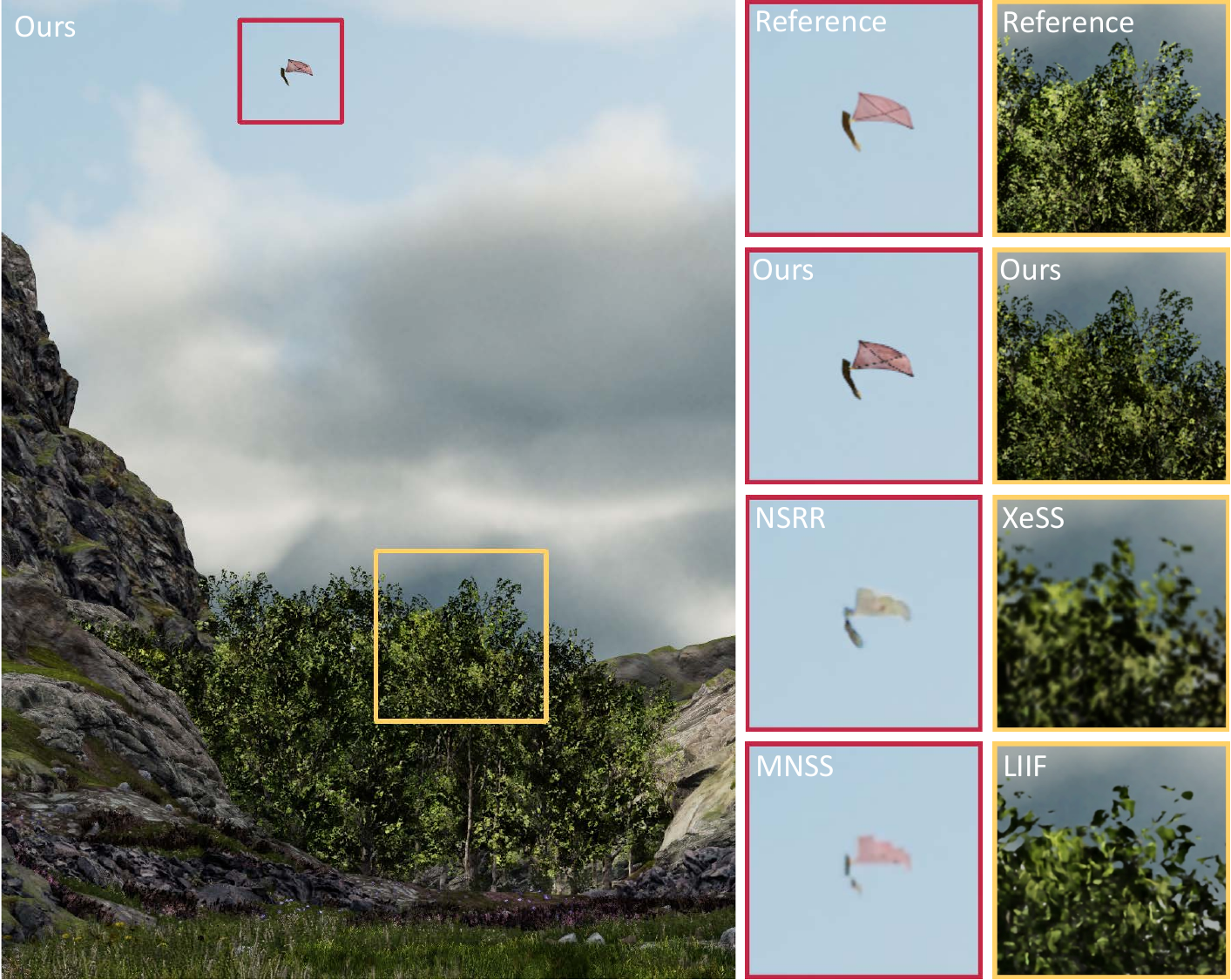} & \includegraphics[width=0.49\linewidth]{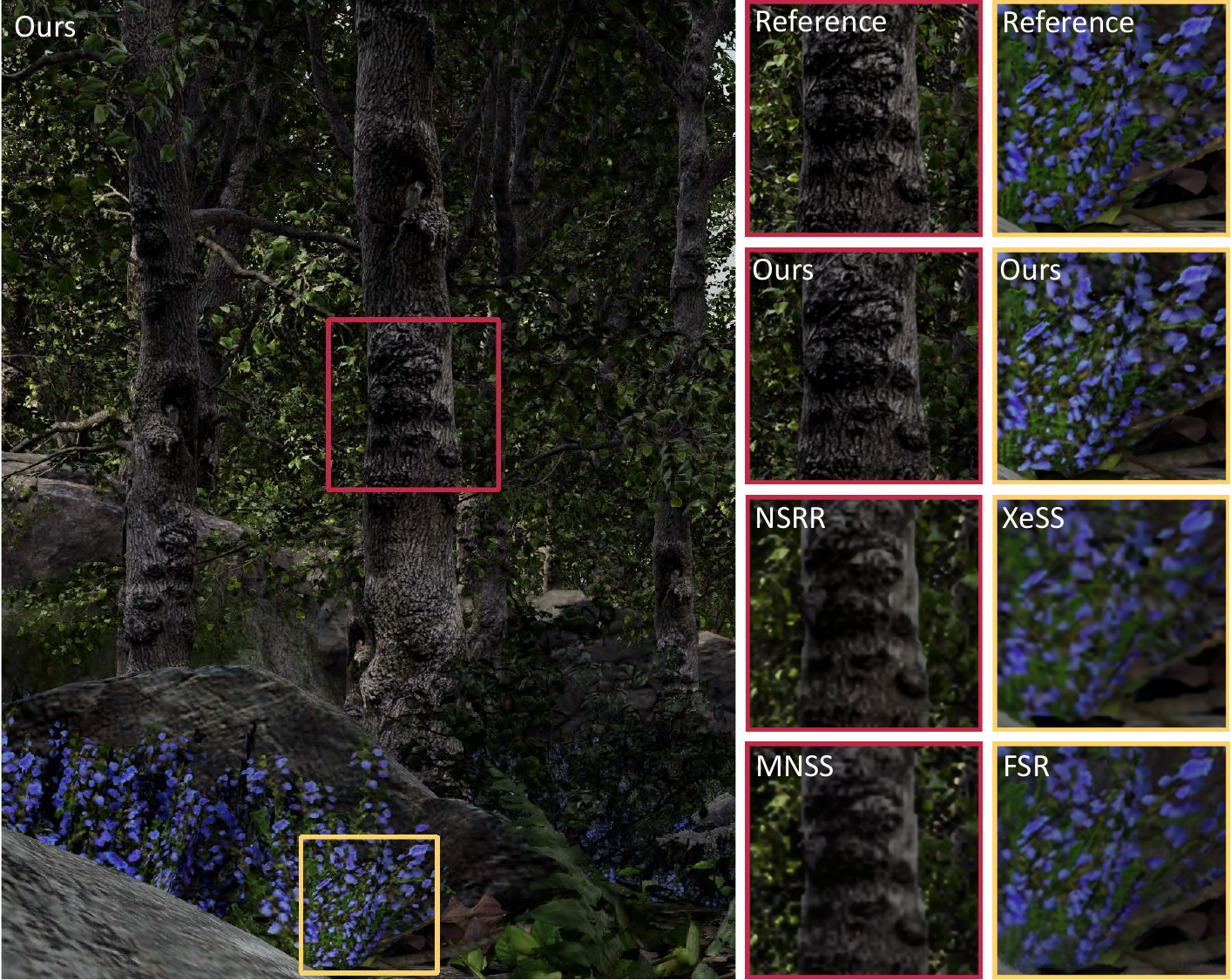} \\
        \includegraphics[width=0.49\linewidth]{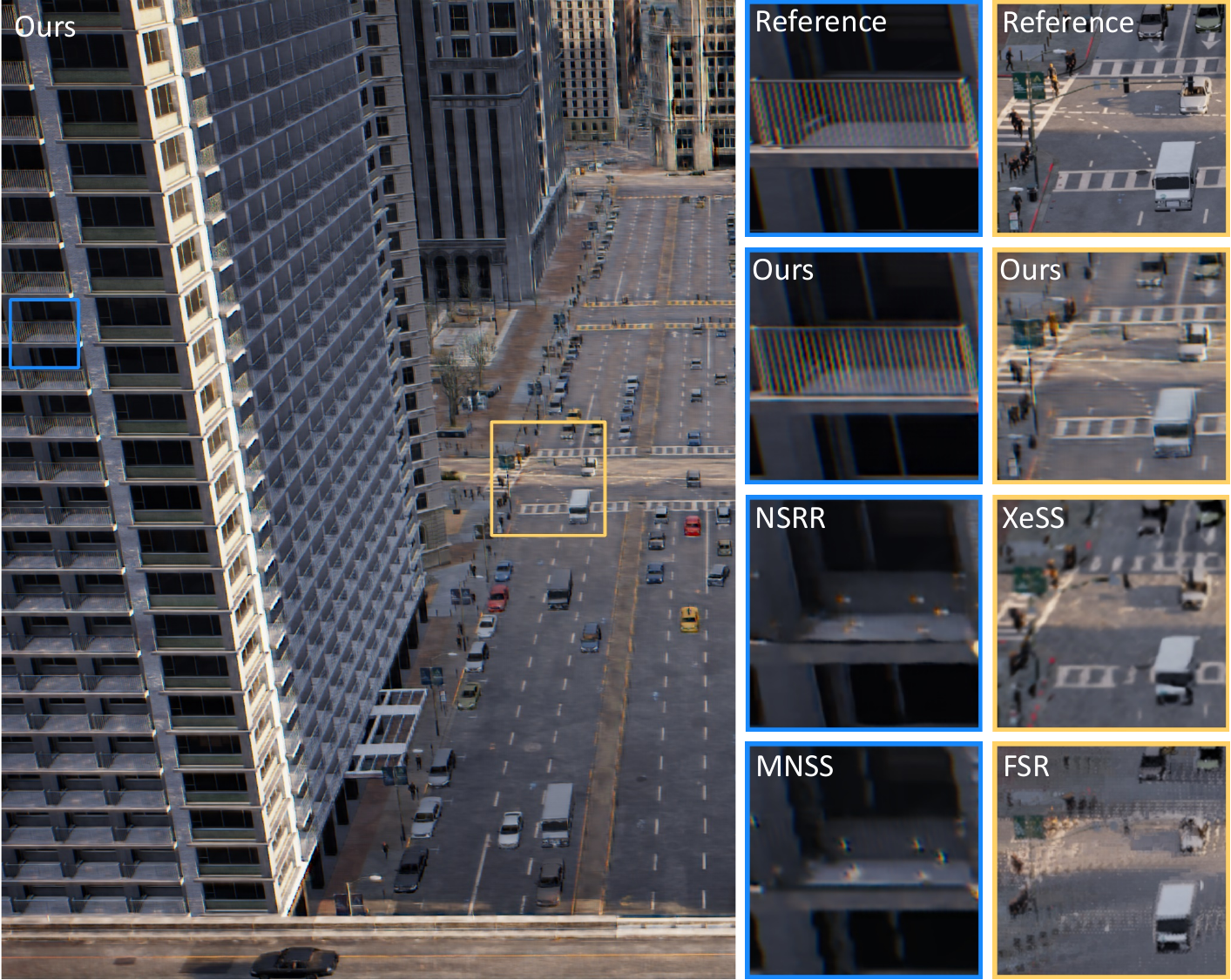} & \includegraphics[width=0.49\linewidth]{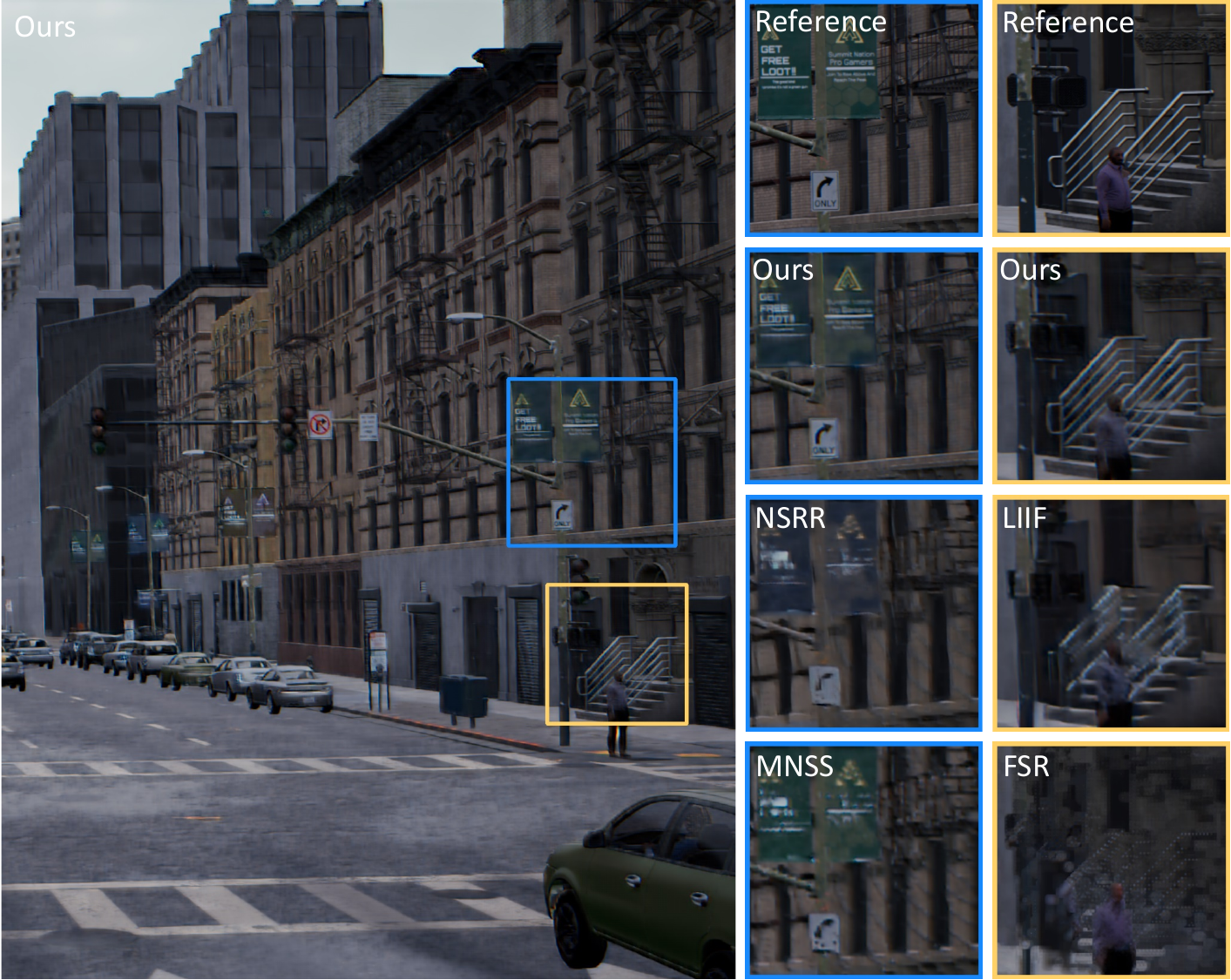} \\
        \includegraphics[width=0.49\linewidth]{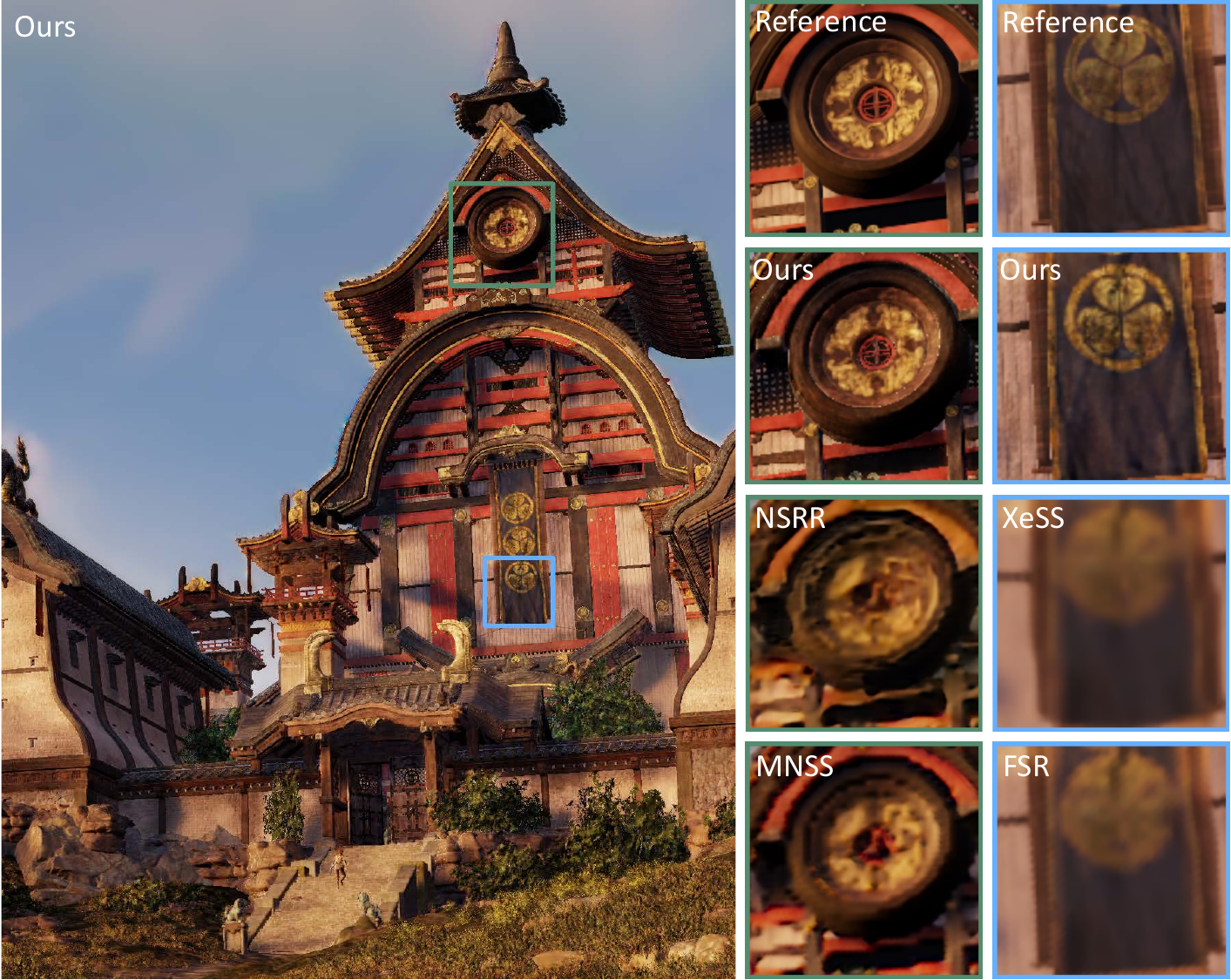} & \includegraphics[width=0.49\linewidth]{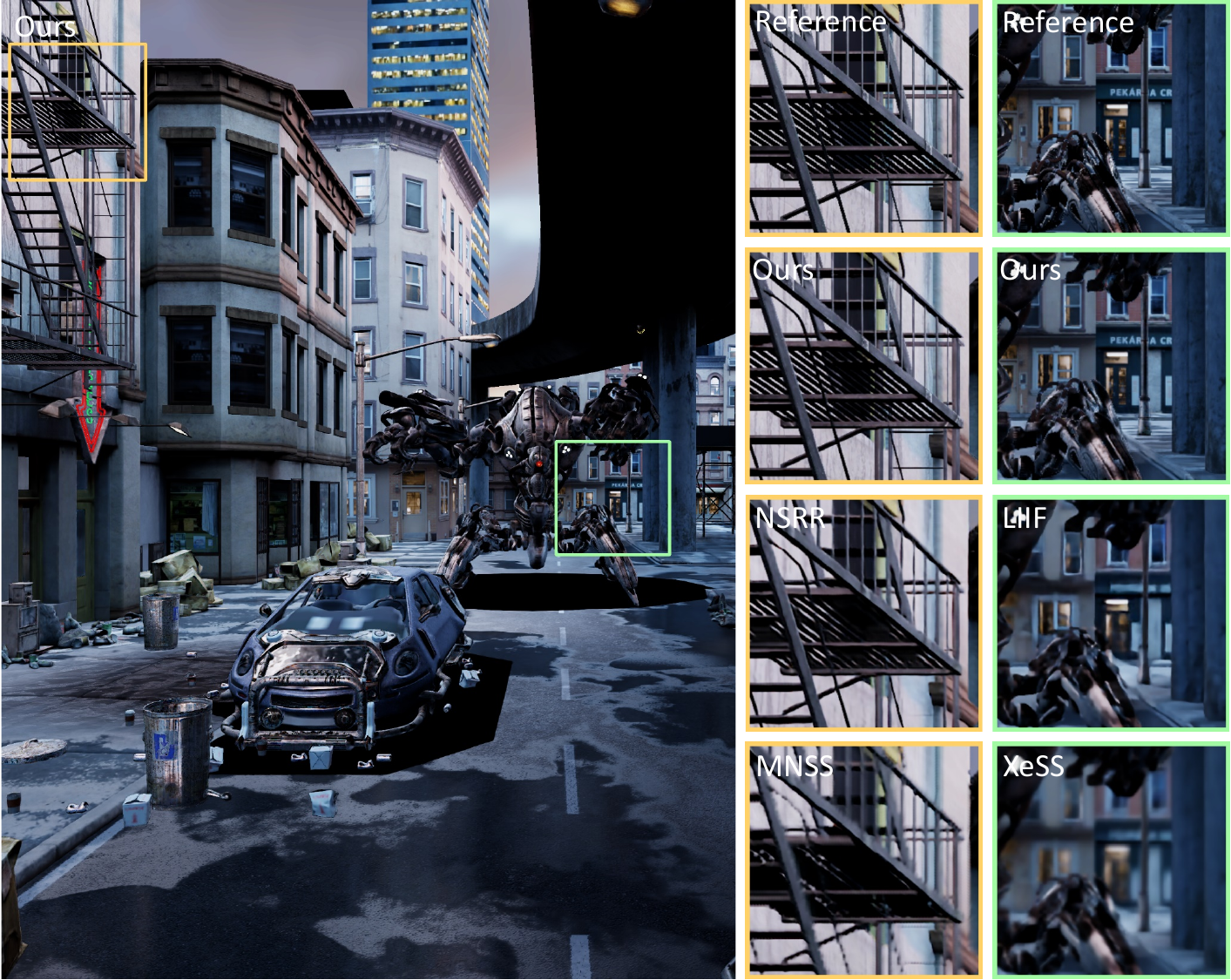}
    \end{tabular}
    \caption{Comparisons of $4\times4$ super-resolution between our method and baseline methods in \texttt{Kite}, \texttt{City}, \texttt{Slay} and \texttt{Showdown} scenes. Full images of reference and baselines are presented in the supplementary material.}
    \label{fig:result_4x}
\end{figure*}

\begin{figure*}[htbp]
    \setlength{\tabcolsep}{2pt}
    \centering
    \includegraphics[width=0.92\linewidth]{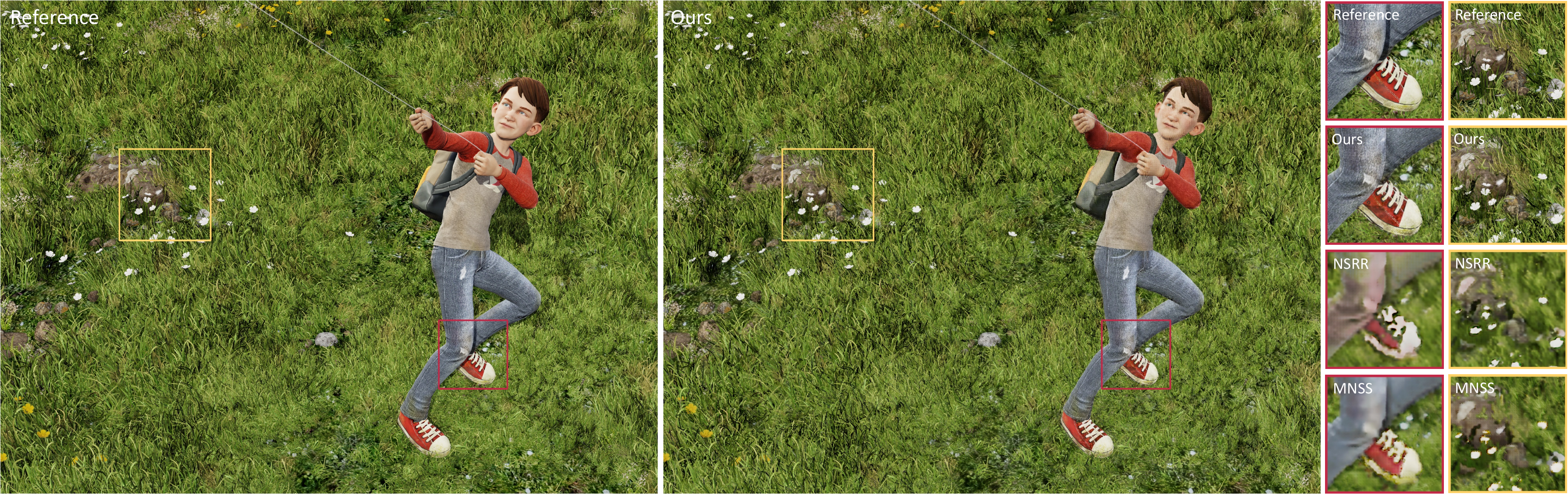} \\
    \includegraphics[width=0.92\linewidth]{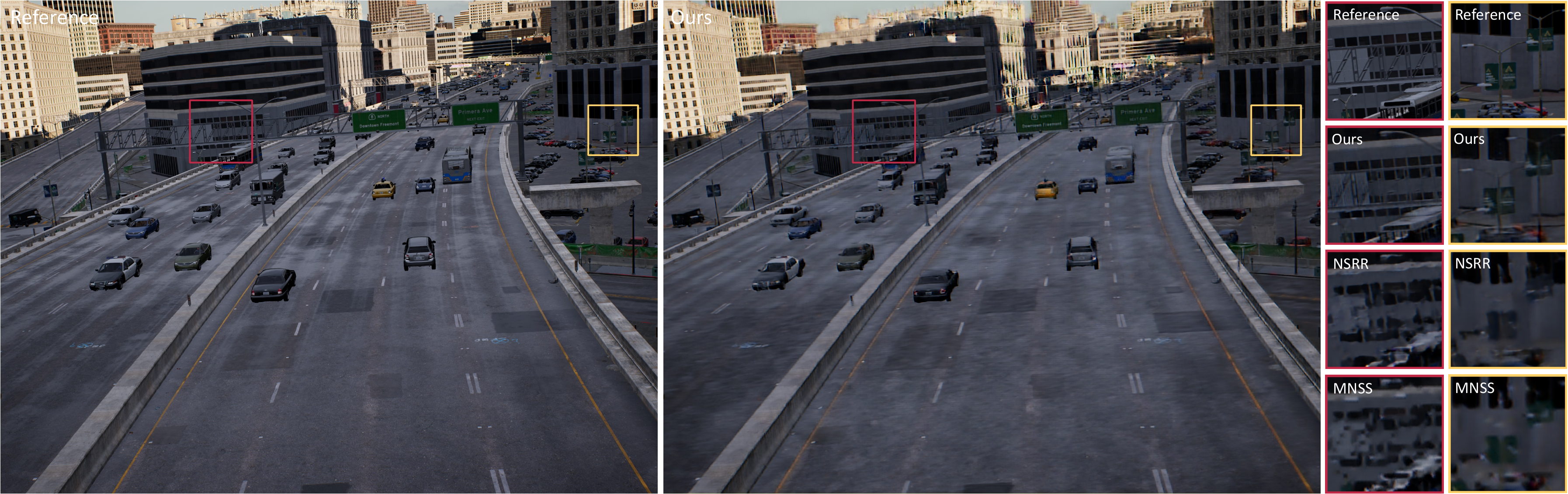} \\
    \includegraphics[width=0.92\linewidth]{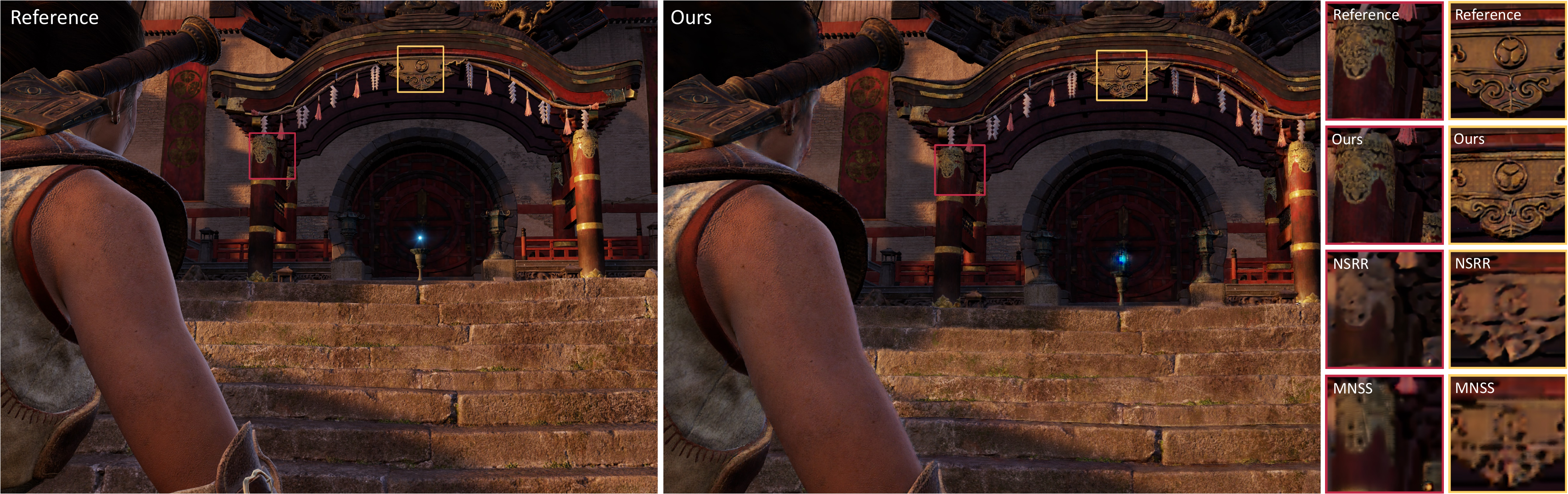} \\
    \includegraphics[width=0.92\linewidth]{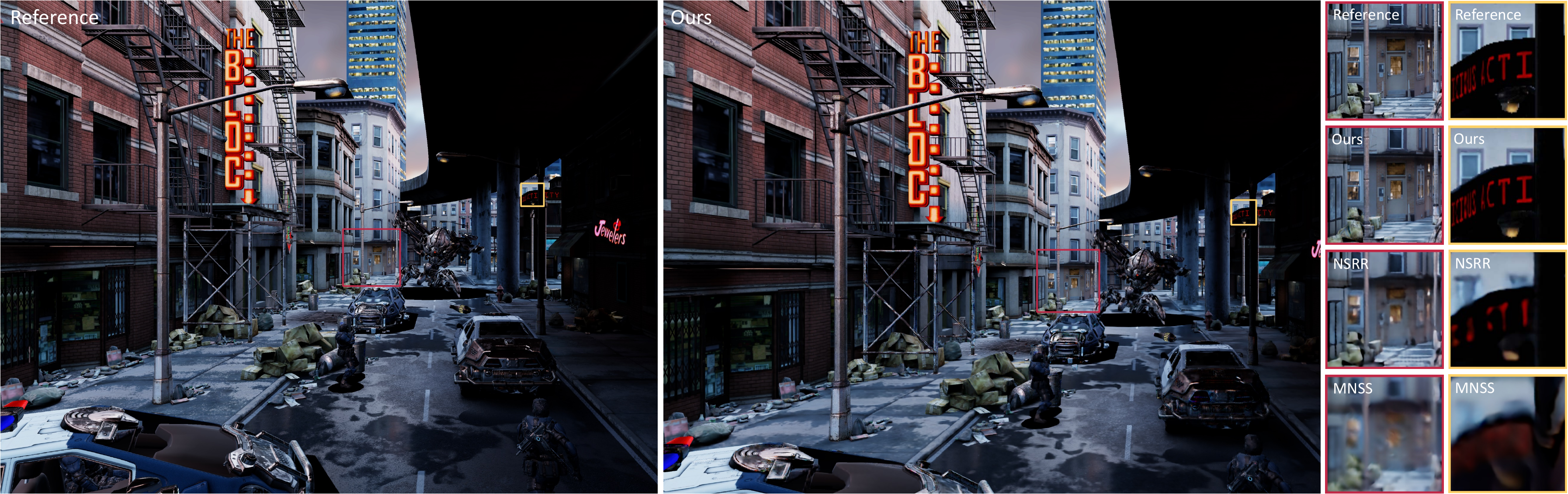}
    \caption{Comparisons of $8\times8$ super-resolution between our method and baseline methods in \texttt{Kite}, \texttt{City}, \texttt{Slay} and \texttt{Showdown} scenes.}
    \label{fig:result_8x}
\end{figure*}

\end{document}